\documentclass[%
 reprint,
 amsmath,amssymb,
 aps,
pra,
superscriptaddress
longbibliography
]{revtex4-2}
\usepackage{amsmath}
\usepackage{color}
\usepackage{graphicx, lipsum} 
\usepackage{natbib}
\usepackage{dcolumn}
\usepackage{bm}
\usepackage{multirow}
\usepackage{comment}
\usepackage{array,booktabs}
\usepackage{xcolor}
\usepackage{hyperref}
\usepackage{mathrsfs}
\usepackage{soul}
\usepackage{tikz}
\usepackage[most]{tcolorbox}
\usetikzlibrary{shapes.geometric, arrows, positioning, calc} 

\usepackage[ruled,vlined,linesnumbered]{algorithm2e}

\begin{document}

\preprint{APS/123-QED}

\title{Alpha helices are more evolutionarily robust to environmental perturbations\\than beta sheets: Bayesian learning and statistical mechanics for protein evolution}



\author{Tomoei Takahashi}
\email{takahashi-tomoei@g.ecc.u-tokyo.ac.jp}
\affiliation{Institute for Physics of Intelligence, Graduate School of Science, The University of Tokyo, 7-3-1 Hongo, Bunkyo-ku, Tokyo 113-0033, Japan.}
\author{George Chikenji}%
\affiliation{%
Graduate School of Engineering, Nagoya University, Furo-cho, Chikusa-ku, Nagoya, 464-8603, Japan.
}%


\author{Kei Tokita}
\affiliation{Graduate School of Informatics, Nagoya University, Furo-cho, Chikusa-ku, Nagoya, 464-8601, Japan.}

\author{Yoshiyuki Kabashima}
\affiliation{Institute for Physics of Intelligence, Graduate School of Science, The University of Tokyo, 7-3-1 Hongo, Bunkyo-ku, Tokyo 113-0033, Japan.}
\affiliation{Department of Physics, Graduate School of Science, The University of Tokyo, 7-3-1 Hongo, Bunkyo-ku, Tokyo 113-0033, Japan}
\affiliation{Trans-Scale Quantum Science Institute, The University of Tokyo, 7-3-1 Hongo, Bunkyo-ku, Tokyo 113-0033, Japan}



\date{\today}

\begin{abstract}
How typical elements that shape organisms, such as protein secondary structures, have evolved, or how evolutionarily susceptible/resistant they are to environmental changes, are significant issues in evolutionary biology, structural biology, and biophysics. According to Darwinian evolution, natural selection and genetic mutations are the primary drivers of biological evolution. However, the concept of ``robustness of the phenotype to environmental perturbations across successive generations," which seems crucial from the perspective of natural selection, has not been formalized or analyzed. In this study, through Bayesian learning and statistical mechanics we formalize the stability of the free energy in the space of amino acid sequences that can design particular protein structure against perturbations of the chemical potential of water surrounding a protein as such robustness. This evolutionary stability is defined as a decreasing function of a quantity analogous to the susceptibility in the statistical mechanics of magnetic bodies specific to the amino acid sequence of a protein. Consequently, in a two-dimensional square lattice protein model composed of 36 residues, we found that as we increase the stability of the free energy against perturbations in environmental conditions, the structural space shows a steep step-like reduction. Furthermore, lattice protein structures with higher stability against perturbations in environmental conditions tend to have a higher proportion of $\alpha$-helices and a lower proportion of $\beta$-sheets. This result is qualitatively confirmed by comparing the histograms of the percentage of secondary structures of evolutionarily robust proteins and randomly selected proteins through an empirical validation using a protein database. The result shows that protein structures rich in $\alpha$-helices are more robust to environmental perturbations through successive generations than those rich in $\beta$-sheets.

\end{abstract}

\maketitle

\section{INTRODUCTION}
\label{sec:intro}
Understanding whether fundamental elements that shape organisms are evolutionarily robust or prone to change is crucial for addressing why the phenotypes of existing organisms are so limited compared to the physically possible patterns, or for resolving significant issues such as predicting evolution. 
According to Darwinian evolution, the two key forces driving evolution are genetic mutations and natural selection. Therefore, when addressing protein evolution, the crucial concepts linked to the aforementioned problems, are the relationships between mutational/environmental robustness and protein structure. In particular, about environmental robustness, natural selection is the phenomenon by which differences in environmental fitness become apparent in subsequent generations. Hence, it is important to consider whether a given protein structure remains robust against environmental changes even in later generations.

Many studies on protein evolution suggest that mutational robustness may have driven the evolution of characteristic protein structures\cite{sikosek2014biophysics, xia2002roles, bloom2007evolution, jayaramen2022mechanisms, vila2024analysis}. It has been shown that there is a correlation between mutational robustness and the geometric symmetry of protein structures\cite{hartling2008mutational}, that secondary structures are more robust to mutations than intrinsic disorder structures\cite{schaefer2010protein}, and that mutational robustness and the structural modularity of proteins (i.e., the proportion of amino acid residues forming secondary structures) contribute to the evolvability of proteins\cite{rorick2010structural, rorick2011protein}. Furthermore, the low algorithmic complexity of gene sequences achieves symmetric protein structures\cite{johnston2022symmetry}. It is also demonstrated that mutational robustness is well-compatible with the functional sensitivity of proteins\cite{tang2020functional}, and there is a correlation between the dynamics of protein structures and their evolvability\cite{tang2021dynamics}. These various lines of research strongly suggest that mutational robustness (and additionally, the low algorithmic complexity of genes) drives the formation of protein secondary structures and the evolvability of proteins.

In addressing the evolution problem, a statistical mechanics approach to abstract models has also produced significant results. A study using the spin glass model has shown that, in evolved organisms, plasticity due to environmental fluctuations and plasticity due to mutations are strongly correlated, leading to a dimensional reduction in the phenotypic space\cite{sakata2020dimensional}. Studies on gene regulatory network (GRN) models have revealed that GRNs with high fitness exhibit high mutational robustness. This tendency is stronger in GRNs obtained through evolutionary simulations than those obtained through efficient sampling methods for exploring high-fitness GRNs\cite{nagata2020emergence, kaneko2022evolution}. It has also been found that both mutational robustness and developmental robustness drive the evolution of GRNs\cite{kaneko2007evolution, ciliberti2007robustness}. These studies suggest that biological evolution possesses mechanisms different from typical optimization processes, enabling the selection of GRNs (phenotypes) with high mutational robustness and noise during development. These findings imply the reduction of phenotypic space.

The above results have significantly advanced our understanding of the relationship between mutational robustness, developmental robustness, and evolvability in proteins (or life in general). However, these concepts of robustness pertain solely to the stability of an individual (or phenotype) against various perturbations over its lifetime. Since evolution involves changes in genetic information and associated traits over generations, considering the stability in terms of how well a trait (phenotype) adapts to or is maladapted to its environment, how this influences the genotype that produces the phenotype, and how these influences alter traits in subsequent generations, helps understand evolution.

In this study, we define the free energy of the space of amino acid sequences (i.e., genetic information) that can design a particular protein structure, utilizing the framework of Bayesian learning. We discuss the stability of this free energy against perturbations in environmental conditions surrounding the protein. The stability of the free energy is determined for a randomly generated two-dimensional (2D) lattice protein model, and we elucidate the relationship between the structural features of the protein and the stability of its free energy.

We use lattice proteins for model proteins\cite{lau1989lattice}. Random structural patterns that have not evolved do not exist in protein structure databases such as the Protein Data Bank (PDB)\cite{berman2000protein}. Therefore, artificial models like lattice proteins are more suitable for this study. We show the definition of secondary structures in 2D lattice proteins of this study in \ref{subsec:definition}. In the analysis of secondary structures of 2D lattice proteins, $\beta$-sheets decrease as the designability (the number of amino acid sequences that fold into a given protein structure) increases\cite{chen2001secondary}. Lattice proteins are also effective for analyzing the free energy landscape of protein folding\cite{cieplak2013energy, shi2016characterizing}, the phenomenon of cold denaturation where proteins denature at low temperatures\cite{van2016consistent}, and the impact of amino acid residue mutations on the native structure of proteins\cite{holzgrafe2011mutation, shi2014effect}. Additionally, lattice models are used to analyze the folding energy landscape of RNA\cite{chen2000rna}. Therefore, if one develops a valid theory for protein evolution, it could be said that lattice proteins can reveal qualitatively accurate behaviors in analyzing the environmental robustness of protein secondary structures across successive generations, our objective in this study.

\section{MODEL AND METHOD}
\subsection{Hamiltonian of lattice HP model with the water chemical potential}

The lattice HP model places amino acids at lattice points, representing the protein structure as a self-avoiding walk on the lattice. A self-avoiding walk is a path that does not pass through the same point more than once on a lattice (or graph). The naturally occurring 20 types of amino acids are simplified into two types: hydrophobic (H), which repels water molecules, and polar (P), which attracts water molecules.

The structure of a protein, denoted as $\bm R$, is represented by the set of coordinates $\bm r_i$ for each amino acid. For a protein with $N$ amino acids, $\bm R = {\bm r_{1}, \bm r_{2}, \cdots, \bm r_{N} }$. The state of the $i$-th amino acid is denoted by $\sigma_{i}$, with $\sigma_{i} = 1$ (for hydrophobic) and $0$ (for polar). The lattice HP model typically considers only the attractive interactions between hydrophobic amino acids. However, since proteins also interact with water molecules surrounding them, in this study, the Hamiltonian of a protein with structure $\bm R$ and sequence $\bm \sigma$ is expressed as follows, with $\mu$ representing the chemical potential of water near the protein surface:
\begin{eqnarray} \label{Hamiltonian} H(\bm R , \bm \sigma; \mu) = -\sum_{i<j}\sigma_{i}\sigma_{j}\Delta(\bm r_{i} - \bm r_{j}) - \mu \sum_{i = 1}^{N} (1 - \sigma_{i}), \end{eqnarray}
where $\Delta(\bm r_{i} - \bm r_{j})$ is a ``contact function" that equals 1 when the $i$-th and $j$-th amino acids are spatially nearest neighbors but not consecutive in the sequence, and 0 otherwise. The second term represents the interaction of polar amino acids with surrounding water molecules. 

This Hamiltonian, proposed for the first time in our previous study\cite{takahashi2021lattice}, includes a hydration term, as shown in Eq. (\ref{Hamiltonian}), which is crucial for exploring the environmental robustness of protein structures\cite{bianco2017role}. Additionally, Eq. (\ref{Hamiltonian}) can be seen as a simplified version of the protein energy in water, excluding interactions between water around the protein and bulk water, as shown in studies like\cite{lazaridis1999effective}. It is important to note that $\mu$ is not a parameter representing the overall environment within an organism but rather the environment surrounding a specific protein.

The chemical potential of water surrounding a protein, $\mu$, can be considered a parameter representing environmental conditions, as it depends on the state of bulk water (such as pH and pressure) and temperature. The strength of hydrogen bonds between hydrophilic amino acid residues and water molecules is generally influenced by temperature. Consequently, $\mu$ can be regarded as an environmental condition specific to the protein structure.

\subsection{Bayesian learning framework}
Bayesian learning is a framework for machine learning based on Bayesian statistics, in which one updates the prior probability (prior) of an event to a posterior probability (posterior) in light of observed data. Statistical models derived from Bayesian learning are often interpreted as probabilistic generative models of observational data. We utilize the properties of Bayesian learning to construct a probabilistic generative model for the phenotype of proteins, namely their native structures. The probabilistic generative model of protein structures described below was devised in our previous work as a method for protein design and has achieved a certain level of success in the context of lattice protein models\cite{takahashi2021lattice, takahashi2022cavity}. Protein design problem is the inverse problem of protein structure prediction. Protein design is thus determining the sequence of amino acids that will fold into a given protein structure \cite{coluzza2017computational, cocco2018inverse}.

In our model, we consider the native structure $\bm R$ as the observed variable, the amino acid sequence $\bm \sigma$ as the latent variable, and the chemical potential of the surrounding water $\mu$ as the hyperparameter. In this context, Bayes' theorem is expressed as follows:
\begin{align}
\label{Bayes Theorem}
    p(\bm \sigma | \bm R, \mu) = \frac{p(\bm R|\bm \sigma, \mu) p(\bm \sigma|\mu)}{\sum_{\bm \sigma}p(\bm R|\bm \sigma, \mu) p(\bm \sigma|\mu)}
\end{align}

Let $\beta$ the inverse temperature of the environment, the likelihood function $p(\bm R|\bm \sigma)$, prior $p(\bm \sigma|\mu)$, and posterior $p(\bm \sigma | \bm R, \mu)$ are respectively given as
\begin{align}
\label{likelihood function}
     p(\bm{R} | \bm \sigma, \mu) =& \frac{e^{-\beta H(\bm{R} , \bm{\sigma} ; \mu)}}{Z(\bm \sigma ; \beta, \mu)},\\
\label{prior}
     p(\bm \sigma | \mu) =& \frac{Z(\bm \sigma; \beta_{p}, \mu_{p})}{\Xi(\beta_{p}, \mu_{p})},\\
    \label{posterior}
    p(\bm \sigma|\bm R, \mu) =& \frac{e^{-\beta H(\bm{R}, \bm{\sigma} ; \mu) } }{Y(\bm R;\beta, \mu)}.
\end{align}
The normalization constants of Eqs. (\ref{likelihood function}), (\ref{prior}), and (\ref{posterior}) are as follows:
\begin{align}
\label{structural partition function}
    Z(\bm \sigma ; \beta, \mu) &= \sum_{\bm R} e^{-\beta H(\bm{R} , \bm{\sigma} ; \mu)},\\
    \label{grand partition function}
    \Xi(\beta_{p}, \mu_{p}) &= \sum_{\bm \sigma} \sum_{\bm R} e ^ {- \beta_{p} H(\bm{R}, \bm{\sigma}; \mu_{p})},\\
    \label{sequence partition function}
    Y(\bm R;\beta, \mu) &= \sum_{\bm \sigma} e^{-\beta H(\bm{R}, \bm{\sigma} ; \mu)}.
\end{align}
We refer to Eqs. (\ref{structural partition function}), (\ref{grand partition function}), and (\ref{sequence partition function}) as the structural partition function, grand partition function, and sequence partition function, respectively. Additionally, the inverse temperature and chemical potential in the prior Eq. (\ref{prior}) may differ from those of protein folding and are thus denoted as $\beta_{p}$ and $\mu_{p}$, respectively.

An important point is that the posterior Eq.(\ref{posterior}) does not include the structural partition function $Z(\bm \sigma ; \beta, \mu)$. This is because $Z(\bm \sigma ; \beta, \mu)$ requires an exhaustive structural search of the Boltzmann factor, which is infeasible considering the infinite degrees of freedom of protein structures. The sum over sequences $\sum_{\bm \sigma}$ is much more manageable than the sum over structures $\sum_{\bm R}$.

The likelihood function $p(\bm R | \bm \sigma, \mu)$ represents the probability of a given structure $\bm R$ occurring for a given amino acid sequence $\bm \sigma$. This is the probability that $\bm \sigma$ folds into structure $\bm R$. More generally, it is the probability that a given genotype results in a given phenotype. Eq. (\ref{likelihood function}) asserts that the likelihood function $p(\bm R | \bm \sigma, \mu)$ is the Boltzmann distribution of structure $\bm R$ conditional on the amino acid sequence $\bm \sigma$. This setting of the likelihood functions $p(\bm R | \bm \sigma, \mu)$ is based on Anfinsen's dogma\cite{anfinsen1973principles}, which states that the state in which a protein adopts its native structure is a thermodynamic equilibrium determined by its amino acid sequence under physiological conditions.

The prior in Eq. (\ref{prior}) is highly non-trivial. To briefly explain the background of our setting of the prior in Eq. (\ref{prior}), it is based on the free energy of a protein with amino acid sequence $\bm \sigma$,
\begin{eqnarray}
\label{Free energy of a sequence}
    F(\bm \sigma; \beta_{p}, \mu_{p}) = -\frac{1}{\beta_{p}} \log Z(\bm \sigma; \beta_{p}, \mu_{p})
\end{eqnarray}
which is proportional to the structural partition function $Z(\bm \sigma ; \beta_{p}, \mu_{p})$ included in it. Since low free energy implies a large partition function, the prior in Eq. (\ref{prior}) can be interpreted as the hypothesis that "amino acid sequences with lower free energy under specific temperature $\beta_{p}$ and chemical potential $\mu_{p}$ have evolved." We call this the hypothesis of sequence weights (HSW). HSW was first proposed in \cite{takahashi2021lattice}. Amino acid sequences with high free energy Eq.(\ref{Free energy of a sequence}) tend to increase the Hamiltonian Eq. (\ref{Hamiltonian}) for many structures. The probability that such amino acid sequences form specific compact three-dimensional structures is extremely low. HSW is a hypothesis that preemptively excludes such sequences.

HSW is still an unverified hypothesis, but our previous studies\cite{takahashi2021lattice, takahashi2022cavity} have shown that a protein design method assuming HSW exhibits high performance for the 2D lattice HP model with $N \leq 36$, which we will analyze in this study. Protein design problem is the inverse problem of protein structure prediction. Additionally, using the prior Eq. (\ref{prior}) based on HSW allows us to cancel out the two partition functions $Z(\bm \sigma ; \beta, \mu)$ and $\Xi(\beta, \mu)$ required for deriving the posterior, thereby avoiding the computational explosion associated with structural searches.

If $\beta = \beta_{p}, \mu = \mu_{p}$ holds, the derivation of the posterior $p(\bm \sigma | \bm R, \mu)$ is then,
\begin{align}
            p({\bm \sigma}|\bm R, \mu) &= \frac{p(\bm R|\bm \sigma, \mu)p(\bm \sigma | \mu)}{\sum_{\bm \sigma}p(\bm R|\bm \sigma, \mu)p(\bm \sigma|\mu)}\\
			&=
			\frac{
   \frac{e^{-\beta H(\bm{R} , \bm{\sigma} ; \mu)}}{Z(\bm \sigma; \beta, \mu)}
			\cdot
			\frac{Z(\bm \sigma;\beta, \mu)}{\Xi(\beta, \mu)}
                    }
                    {
			\sum_{\bm \sigma}
			\frac{e^{-\beta H(\bm{R} , \bm{\sigma} ; \mu)}}{Z(\bm \sigma;\beta, \mu)}
			\cdot
			\frac{Z(\bm \sigma;\beta \mu)}{\Xi(\beta, \mu)}}\\
   \label{eq14}
            &= \frac{
                    \frac{
                        e^{-\beta H(\bm{R} , \bm{\sigma} ; \mu)}
                        }
                        {
                        \Xi(\beta, \mu)
                        }
                    }
                    {\sum_{\bm \sigma}
                    \frac{
                        e^{-\beta H(\bm{R} , \bm{\sigma} ; \mu)}
                        }
                        {
                        \Xi(\beta, \mu)
                        }
                    }\\
                \label{eq15}    
            &= \frac{e^{-\beta H(\bm{R} , \bm{\sigma} ; \mu)}}{\sum_{\bm \sigma} e^{-\beta H(\bm{R} , \bm \sigma;\mu)}}.
\end{align}
From Eq. (\ref{eq14}) to Eq. (\ref{eq15}), We used that the grand partition function $\Xi(\beta, \mu)$ does not depend on the amino acid sequence $\bm \sigma$, allowing it to be factored out of the sum $\sum_{\bm \sigma}$. In the posterior $p(\bm \sigma | \bm R, \mu)$, the sequence space considered can be regarded, from an evolutionary perspective, as the set of typical sequences that realize a given structure $\bm R$.

Eq. (\ref{Free energy of a sequence}) represents the free energy of a protein with a given amino acid sequence $\bm \sigma$. Hence, it is natural to assume that the inverse temperature and chemical potential in this equation are equal to those in the likelihood function Eq. (\ref{likelihood function}), which represents the probability that the protein folds into a given structure $\bm R$.

\subsection{A free energy depends on a protein structure and its stability to the environmental perturbation}
\label{subsec:empirical}

\begin{figure}[tb]
\begin{center}
\vspace{5mm}
\includegraphics[width=1.0\linewidth]{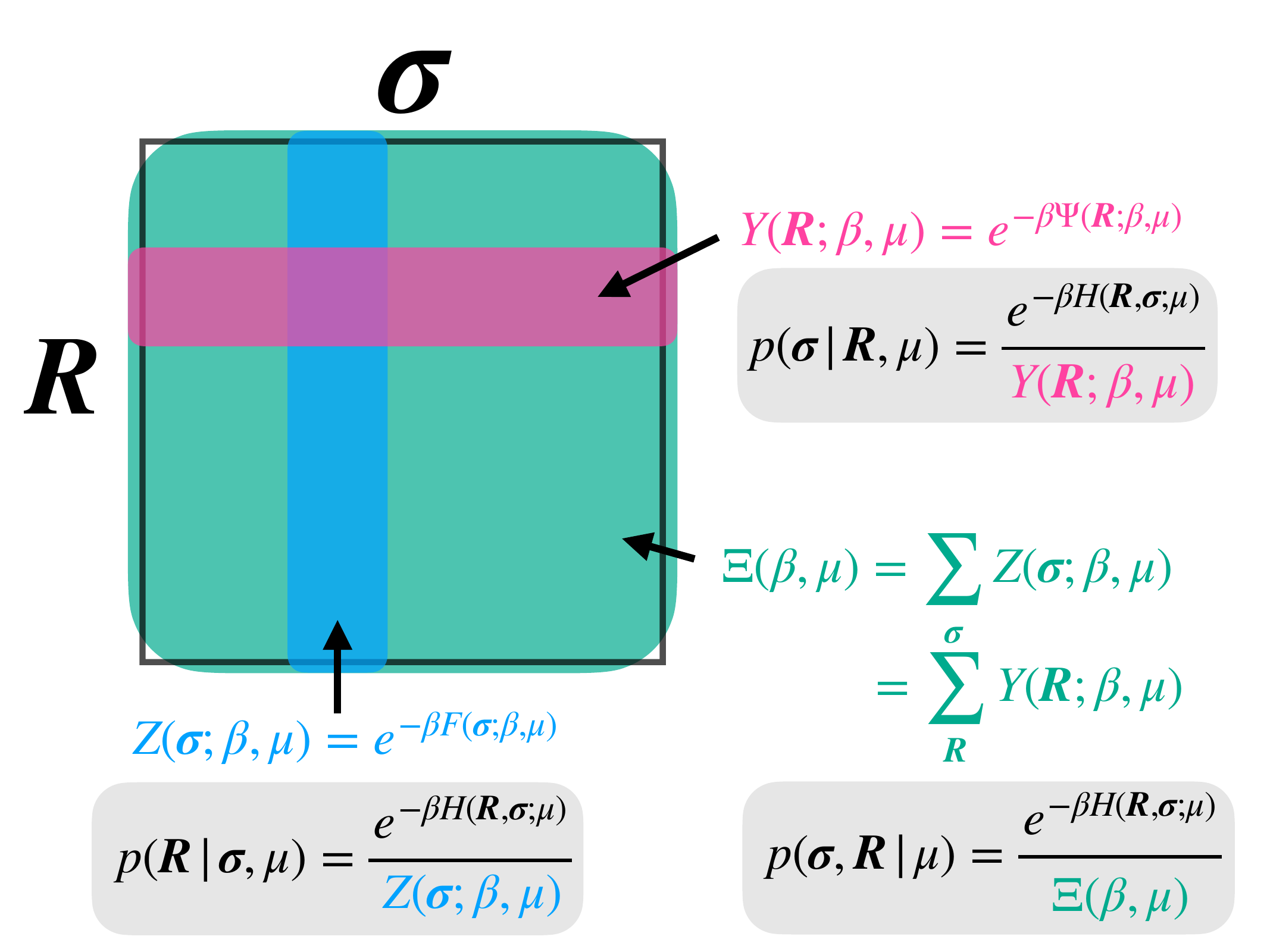}
\end{center}
\caption{A diagram illustrating the conditional probabilities introduced so far, their corresponding partition functions as normalization constants, and the specific quantities computed by these partition functions. The rows represent the degrees of freedom associated with structure $\bm R$ while the columns represent the degrees of freedom associated with sequence 
$\bm \sigma$.
By summing the Boltzmann factor $e^{-\beta H(\bm R,\bm \sigma;\mu)}$ along the rows while keeping a column fixed, the structural partition function (\ref{structural partition function}) is obtained (blue shade). Conversely, summing over the columns while keeping a row fixed yields the sequence partition function (\ref{sequence partition function}) (red shade). Summing over both rows and columns results in the partition function for both structure and sequence \ref{grand partition function} (green shade). Each partition function is represented in the figure using free energy. The probability distributions shown beneath each partition function correspond to the associated conditional probability distributions, where each partition function serves as the normalization constant for the respective probability distribution.
}
\label{fig:Bayes_calc_image}
\end{figure}

In this subsection, we define the free energy for a sequence space dependent on a particular protein structure $\bm R$, and derive expressions demonstrating its stability. To facilitate this, we present the expression for the marginal likelihood $p(\bm R | \mu)$ here. In Bayesian learning, the marginal likelihood represents the probability of observing the data given a value for the hyperparameter. For the lattice protein case under discussion, the marginal likelihood is
\begin{align}
    p(\bm R | \mu) &= \sum_{\bm \sigma}p(\bm R|\bm \sigma; \mu)p(\bm \sigma | \mu) \nonumber \\
	&=
			\sum_{\bm \sigma}
			\frac{e^{-\beta H(\bm{R} , \bm{\sigma} ; \mu)}}{Z(\bm \sigma; \beta, \mu)}
			\cdot
			\frac{Z(\bm \sigma; \beta_{p}, \mu_{p})}{\Xi(\beta_{p}, \mu_{p})} \nonumber \\
   \label{marginal likelihood by sequence partition function}
	&=
			\frac{Y(\bm R; \beta, \mu)}{\sum_{\bm R}Y(\bm R; \beta, \mu)},
\end{align}
where $\beta = \beta_{p}$ and $\mu = \mu_{p}$ hold. From Eq. (\ref{marginal likelihood by sequence partition function}), the marginal likelihood, i.e., the probability of observing a protein structure $\bm R$ given the environmental conditions represented by the chemical potential of the surrounding water $\mu$, is proportional to the sequence partition function $Y(\bm R; \beta, \mu)$. Furthermore, the marginal likelihood expressed in Eq. (\ref{marginal likelihood by sequence partition function}) is identical to the denominator of the posterior distribution $p(\bm \sigma|\bm R, \mu)$ when $\Xi(\beta, \mu)$ is not canceled out in the transition from Eq. (\ref{eq14}) to Eq. (\ref{eq15}). Therefore, the marginal likelihood $p(\bm R | \mu)$ serves as the partition function for the posterior distribution $p(\bm \sigma|\bm R, \mu)$ as described in Eq. (\ref{eq15}). Consequently, we can consider the corresponding free energy as follows:
\begin{align}
    \label{evolutionary free energy}
    \Psi(\bm R, \mu) = -\frac{1}{\beta}\log p(\bm R | \mu).
\end{align}
Here, Fig. \ref{fig:Bayes_calc_image} schematically illustrates the various conditional probabilities introduced thus far, their corresponding partition functions as normalization constants, and the free energies associated with these partition functions, along with their respective computational interpretations.

When a specific structure $\bm R$ is determined at a given inverse temperature $\beta$, the free energy $\Psi(\bm R, \mu)$ becomes a function solely of the environmental conditions $\mu$. When $\Psi(\bm R, \mu)$ is minimized at a particular environmental condition $\mu = \mu_{EB}$—a condition equivalent to maximizing the marginal likelihood, which is referred to as Empirical Bayes estimation in the field of Bayesian learning—the stability of $\Psi(\bm R, \mu)$ around $\mu = \mu_{EB}$ requires
\begin{align}
\label{kyokudai condition}
    \frac{\partial^{2}}{\partial \mu^{2}} \Psi(\bm R, \mu) \bigg|_{\mu=\mu_{EB}} > 0.
\end{align}
Before we manipulate Eq. (\ref{kyokudai condition}), we define the two types of expected values appearing in the transformation of Eq. (\ref{kyokudai condition}). These are the average taken over the posterior distribution when $\beta = \beta_{p}$ and $\mu = \mu_{p}$ hold, and the average taken over the joint distribution, which is the product of the likelihood function and the prior. The joint distribution is given as follows:
\begin{align}
    \label{joint distribution}
    p(\bm R, \bm \sigma | \mu) &= p(\bm R | \bm \sigma, \mu)p(\bm \sigma | \mu) \nonumber\\
                               &= \frac{e^{-\beta H(\bm R, \bm \sigma;\mu)}}{\sum_{\bm R} \sum_{\bm \sigma} e^{-\beta  H(\bm R , \bm \sigma;\mu)}}.
\end{align}
Thus, the joint distribution $p(\bm R, \bm \sigma | \mu)$ forms a Boltzmann distribution where both the structure $\bm R$ and the sequence $\bm \sigma$ serve as thermal variables. For any physical quantity $X(\bm R, \bm \sigma)$ that is a function of the structure $\bm R$ and the sequence $\bm \sigma$, the average taken over the posterior distribution $p(\bm \sigma | \bm R, \mu)$ and the average taken over the joint distribution $p(\bm R, \bm \sigma | \mu)$ are given as
\begin{align}
    \label{posterior average}
    \langle X(\bm R, \bm \sigma) \rangle_{| \bm R} &:= \sum_{\bm \sigma} X(\bm R, \bm \sigma) p({\bm \sigma}|\bm R, \mu) \nonumber \\
    &= \frac{\sum_{\bm \sigma} X(\bm R, \bm \sigma) e^{-\beta  H(\bm R , \bm \sigma;\mu)}}{\sum_{\bm \sigma} e^{-\beta  H(\bm R , \bm \sigma;\mu)}}, \\
    \langle X(\bm R, \bm \sigma) \rangle &:= \sum_{\bm R} \sum_{\bm \sigma} X(\bm R, \bm \sigma) p(\bm R, \bm \sigma| \mu) \nonumber \nonumber \\
    \label{joint average}
    &= \frac{\sum_{\bm R} \sum_{\bm \sigma} X(\bm R, \bm \sigma) e^{-\beta  H(\bm R , \bm \sigma;\mu)}}{\sum_{\bm R} \sum_{\bm \sigma} e^{-\beta  H(\bm R , \bm \sigma;\mu)}}.
\end{align}
The notation used on the far left side of Eq. (\ref{posterior average}) explicitly indicates that the quantity depends on a specific structure $\bm R$. It is important to note that this does not represent an average over all possible $\bm R$ patterns.

To manipulate inequality (\ref{kyokudai condition}), we substitute Eq. (\ref{marginal likelihood by sequence partition function}) into Eq. (\ref{evolutionary free energy}), and then transform Eq. (\ref{kyokudai condition}) as
\begin{widetext}
\begin{equation}
\begin{aligned}
\label{variance inequality}
    \left < \left(\beta \sum_{i=1}^{N}\left(1-\sigma_{i}\right)\right)^{2}\right>_{| \bm R} - \left(\beta \left< \sum_{i=1}^{N}\left( 1 - \sigma_{i}\right)\right>_{| \bm R}\right)^{2}
    &<
     \left < \left( \beta \sum_{i=1}^{N}\left(1-\sigma_{i}\right)\right)^{2}\right> - \left(\left< \beta\sum_{i=1}^{N}\left( 1 - \sigma_{i}\right)\right>\right)^{2}.\\
\end{aligned}
\end{equation}
\end{widetext}
Thus, Eq. (\ref{kyokudai condition}) results in an inequality between the variance of the number of hydrophilic amino acid residues calculated from the posterior and the variance calculated from the joint distribution. Furthermore, we obtain following expression of Eq. (\ref{variance inequality}):
\begin{align}
\label{eq25}
    \beta \sum_{i,j}\left[ \left < \sigma_{i} \sigma_{j}\right >_{| \bm R} - \left <\sigma_{i} \right >_{| \bm R} \left< \sigma_{j} \right >_{| \bm R}\right]
     <
      \beta \sum_{i,j}\Big[ \left < \sigma_{i} \sigma_{j}\right > - \left <\sigma_{i} \right > \left< \sigma_{j} \right > \Big],
\end{align}
where we denote $\sum_{i,j} = \sum_{i = 1}^{N} \sum_{j=1}^{N}$. The definition of the magnetic susceptibility $\chi$ at the equilibrium state using the spin correlation function is given by $\chi = \beta \sum_{i,j}\left[ \left<\sigma_{i}\sigma_{j}\right> - \left<\sigma_{i} \right> \left<\sigma_{j} \right>\right]$, where $\sigma_{i}$ denotes the $i$-th spin variable, similar to the current context. Thus, under the posterior Eq. (\ref{eq15}), which is the Boltzmann distribution of sequences conditioned on the structure $\bm R$, so to speak the ``hydrophobic susceptibility" can be defined as $\chi_{\bm R} := \beta \sum_{i,j}\left[ \left < \sigma_{i} \sigma_{j}\right >_{| \bm R} - \left <\sigma_{i} \right >_{| \bm R} \left< \sigma_{j} \right >_{| \bm R}\right]$. For the joint distribution, which is the Boltzmann distribution over both spaces of $\bm R$ and $\bm \sigma$, the hydrophobic susceptibility is denoted as $\chi := \beta \sum_{i,j}\Big[ \left < \sigma_{i} \sigma_{j}\right > - \left <\sigma_{i} \right > \left< \sigma_{j} \right > \Big]$. Hence, Eq. (\ref{eq25}) becomes the inequality between those two hydrophobic susceptibilities:
\begin{align}
\label{necessary condition 2}
    \chi_{\bm R} < \chi.
\end{align}
We here consider the evolutionary meaning of $\chi_{\bm R}$. It represents the susceptibility of the hydrophobicity of the structure $\bm R$ to perturbations in $\mu$, within the posterior $p(\bm \sigma | \bm R, \mu)$. Given that changes in the hydrophobic/hydrophilic composition (hydrophobicity) of amino acid residues can lead to alterations in protein structures \cite{bigelow1967average}, $\chi_{\bm R}$ reflects the structural plasticity of protein $\bm R$ to environmental changes via macroscopic shifts in the gene space that facilitates the formation of $\bm R$. Therefore, we propose to call $\chi_{\bm R}$ the evolutionary structural plasticity of a protein structure $\bm R$. The term evolutionary structural plasticity is used to distinguish it from phenotypic plasticity, which refers to an individual's capacity to adapt to environmental changes.

This quantity, $\chi_{\bm R}$, represents the plasticity of a genotype associated with a specific structure $\bm R$ to change for subsequent generations. This concept can be understood by following the simple dynamics of Darwinian evolution. Consider a population of organisms where proteins with the structure $\bm R$ exist in a particular generation. In the subsequent generations, assume that amino acid sequences with slight differences in hydrophobic/hydrophilic composition, which incidentally fold into the same structure $\bm R$, existed either by chance or due to mutations. Subsequently, perturbations in environmental condition $\mu$ occur, leading to the evolution of amino acid sequences that adapt to this new environment, consequently inducing changes in the original structure $\bm R$. If $\chi_{\bm R}$ is high, it indicates a higher such genotypic change; conversely, a low $\chi_{\bm R}$ suggests that changes are less likely. Therefore, $\chi_{\bm R}$ does not merely reflect the susceptibility of the structure $\bm R$ to change within a single generation but rather its plasticity for change in an evolutionary context.

While there exist proteins whose structures remain unchanged despite variations in hydrophobicity, such proteins are a minority among all proteins. Moreover, it can be assumed that even these exceptional proteins would alter their structures if there were significant differences in hydrophobicity.

$\chi$ represents the susceptibility of sequences to change under the joint distribution $p(\bm R, \bm \sigma | \mu)$; thus, $\chi_{\bm R}$ is akin to an average of overall structural patterns. Therefore, Eq. (\ref{necessary condition 2}) asserts that for the free energy $\Psi(\bm R, \mu)$ of a given structure $\bm R$ to be stable against perturbations in $\mu$, the $\chi_{\bm R}$ of that structure must be lower than the average $\chi$ across all structures.

Finally, for structures that satisfy Eq. (\ref{necessary condition 2}), we define the following quantity as the steepness of the function around the minimum of the free energy $\Psi(\bm R, \mu)$, specifically the second derivative with respect to $\mu$: $\frac{\partial^{2}}{\partial \mu^{2}} F(\bm R,\mu) |_{\mu=\mu_{EB}} = \chi - \chi_{\bm R}$,
\begin{align}
\label{environmental robustness}
    \kappa_{\bm R} =  \chi - \chi_{\bm R}.
\end{align}
This quantity $\kappa_{\bm R}$ represents the stability against perturbations in $\mu$ around the minimum state of the free energy $\Psi(\bm R, \mu)$, independent of temperature. 

The evolutionary significance of $\kappa_{\bm R}$ is not immediately apparent. However, since $\chi_{\bm R}$ represents the evolutionary plasticity in response to environmental changes, $\kappa_{\bm R}$, as a decreasing function of $\chi_{\bm R}$, indicates the robustness of $\bm R$ after several generations under environmental perturbations.

It is important to note that even if hydrophobicity remains constant, different amino acid sequences can lead to different structures. Therefore, $\kappa_{\bm R}$ precisely measures the stability against structural changes due to differences in hydrophobicity, which implies a tolerance for macroscopic structural changes while permitting minor structural variations. The ``minor" changes in protein structure that are considered here might include slight alterations due to microscopic differences in the contact network, for example.

For a given structure $\bm R$, the computation of $\kappa_{\bm R}$ are performed using Belief Propagation (BP) (theoretical details are in Appendix \ref{sec:appendixA}). BP is an algorithm that efficiently computes the marginal probability of an element in a graph with complex interactions for the entire probability distribution of the graph. The BP algorithm is derived using an extended method of mean-field approximation called the cavity method. It has been shown that the solutions produced by BP are equivalent to the Bethe approximation\cite{kabashima1998belief}, and if the graph is a tree, the solution is exact. BP also provides good approximate solutions when the graph is close to a tree. Since the contact network of a 2D lattice protein is often a tree graph, BP is suitable for 2D lattice protein models. Indeed, our previous research has shown that the design accuracy of 2D lattice proteins is nearly the same when using BP and Markov Chain Monte Carlo (MCMC) methods\cite{takahashi2022cavity}.

\subsection{Definition of the secondary structure of the\\2D lattice proteins}
\label{subsec:definition}

\begin{figure}[tb]
\begin{center}
\vspace{5mm}
\includegraphics[width=1.0\linewidth]{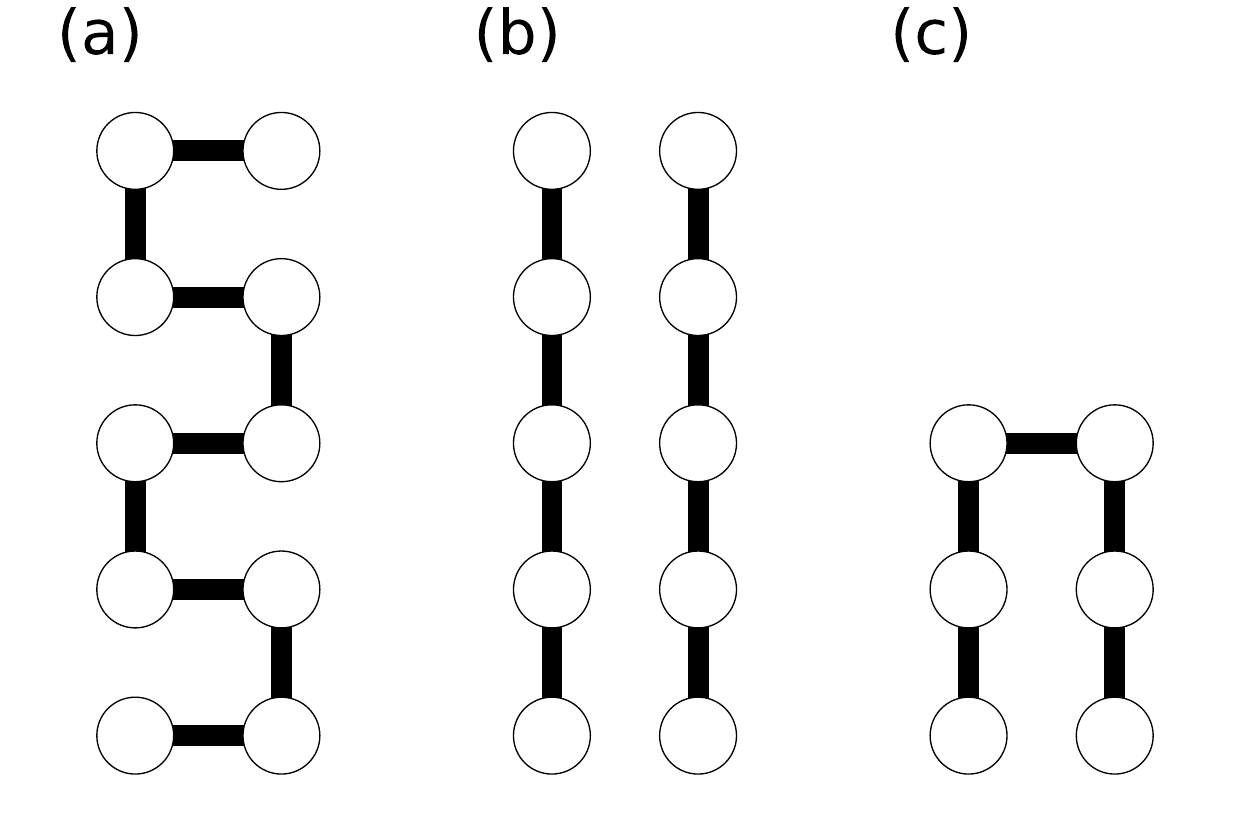} \end{center}
\caption{Examples of secondary structures in lattice proteins used in this study. Both $\alpha$-helices and $\beta$-sheets consist of at least six residues, excluding the anti-parallel sheet structure of 4 residues within a 6-residue $\beta$-turn. (a) Example of a 10-residue $\alpha$-helix extending in the y-axis direction. Structures extending in the y-axis direction are counted separately from their mirror images, while structures extending in the x-axis direction are counted separately from their 90-degree rotated versions. (b) Example of a 10-residue $\beta$-sheet. Parallel and anti-parallel configurations are counted separately, as are structures rotated by 90 degrees in the x-axis direction. (c) Example of a 6-residue $\beta$-turn. The 4-residue anti-parallel sheet structure within the turn (the bottom half of (c)) is counted as a 4-residue $\beta$-sheet. Mirror images and rotations are counted separately.} \label{secondary_structure_example}
\end{figure}

Here we define the secondary structures in 2D lattice proteins, namely $\alpha$-helices and $\beta$-sheets. An $\alpha$-helix is defined as a structure where amino acid residues form a U-shaped structure connected in alternating orientations and a $\beta$-sheet is defined as a structure where amino acid residues are aligned in parallel or anti-parallel configurations (Fig. \ref{secondary_structure_example}). The detection of these secondary structures was performed from contact maps, using the identical method proposed in the previous study \cite{chan1989compact}.

Both $\alpha$-helices and $\beta$-sheets must have at least six residues. This requirement is due to the overrepresentation of single U-shaped structures (four residues) and paired residues in 2D lattice proteins, which would otherwise not qualify as secondary structures. However, a specific case of a 4-residue $\beta$-sheet within a 6-residue $\beta$-turn (Fig. \ref{secondary_structure_example}c) is included. We count $\alpha$-helices and $\beta$-sheets according to these rules, treating all other regions as random coils.

In lattice proteins, $\alpha$-helices and $\beta$-sheets may share one or two amino acid residues. This study prioritizes $\beta$-sheets when two residues are shared and $\alpha$-helices when one residue is shared. If two residues are shared, the $\alpha$-helix is resized by removing the shared residues, and if the resized helix has four residues, it is classified as a random coil. If one residue is shared, the $\beta$-sheet is resized by removing the shared residue and its paired residue, and if the resized sheet has four residues (unless it is part of a 6-residue $\beta$-turn), it is classified as a random coil. Prioritizing $\beta$-sheets for two shared residues and $\alpha$-helices for one shared residue helps balance the two structures.

We do not use 3D lattice proteins because the $\alpha$-helix in three dimensions requires six residues per turn, unlike the real protein case of 3.6 residues per turn\cite{nguyen2009orientation}. In two dimensions, as shown in Fig. \ref{secondary_structure_example}(a), four residues per turn are closer to realistic proteins. This distinction is crucial as we define the proportion of secondary structures in terms of the number of residues forming these structures within a given protein.

In this study, we use maximally compact structures of $N = 6 \times 6$ residues. The total number of structural patterns is 28,732, excluding structures symmetric under $90^{\circ}$ rotation, mirror images along horizontal and vertical axes, diagonal reflections, and head-tail symmetries of the self-avoiding walk.

\section{RESULTS}
\subsection{Changes in the Number of Lattice Protein Structures According to Evolutionary Stability}
\label{subsec:changesinnumber}

\begin{figure}[tb]
\begin{flushleft}
\vspace{5mm}
\includegraphics[width=1.0\linewidth]{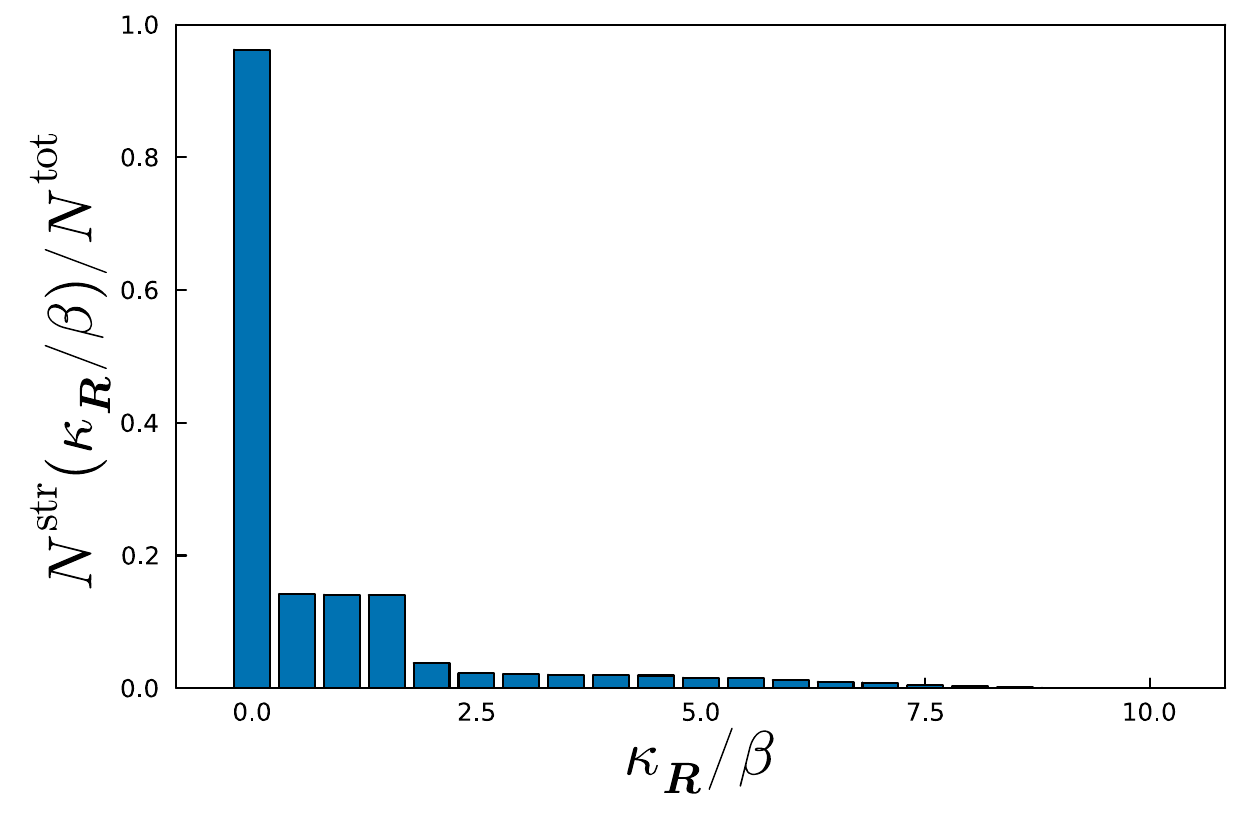}
\end{flushleft}
\caption{We plot a bar graph to visualize the changes in the proportion of structures with stability $\kappa_{\bm R}/\beta$ greater than or equal to each value of $\kappa_{\bm R}/\beta$. Each bar represents the proportion of structures that have a $\kappa_{\bm R}/\beta$ greater than or equal to the corresponding value on the horizontal axis, incrementing $\kappa_{\bm R}/\beta$ from 0 to 10 in steps of 0.5. The total number of maximally compact structures for an amino acid residue count $N = 6 \times 6$ is $N^{tot} = 28732$. The inverse temperature is set at $\beta = 10$. The proportion of structures with $\kappa_{\bm R}/\beta < 0$ is approximately 4\%, indicating a low number of such structures. Additionally, the number of structures with $\kappa_{\bm R}/\beta \geq 9.5$ is zero, as the maximum value of $\kappa_{\bm R}/\beta$ recorded is 9.3971. Readers should note that each bar in the graph represents the proportion of all structures with a $\kappa_{\bm R}/\beta$ greater than or equal to the value at the left end of that bar's range, not just within the range itself.}
\label{6_by_6_changes_of_num_of_struct_with_chi}
\end{figure}

First, we examine how the number of lattice protein structures decreases as the stability $\kappa_{\bm R}$ increases, essentially observing how the dimension of the phenotypic space reduces with changes in $\kappa_{\bm R}$. Specifically, we increment $\kappa_{\bm R}$ from 0 in small steps and record the number of structures that have a $\kappa_{\bm R}$ larger than the value at each step, denoting this number as $N^{\rm str}(\kappa_{\bm R})$. We show the results for the quantity $\kappa_{\bm R}$ divided by $\beta$. This measure is taken to cancel out the $\beta$ in the coefficients of $\chi_{\bm R}$ and $\chi$, which are common across all structures. We then calculate its proportion out of the total number of structures, which is 28,732. The results, showing the reduction in the number of structures with increasing $\kappa_{\bm R}/\beta$ within the range $0 \leq \kappa_{\bm R}/\beta \leq 10$, are displayed in Fig.\ref{6_by_6_changes_of_num_of_struct_with_chi}. In Fig.\ref{6_by_6_changes_of_num_of_struct_with_chi}, $N^{tot}$ represents the total number of maximally compact structures for an amino acid residue number $N = 6 \times 6$, equaling 28,732. We set at $\beta =10$.

From Fig. \ref{6_by_6_changes_of_num_of_struct_with_chi}, it is evident that $N^{\rm str}(\kappa_{\bm R}/\beta)$ decreases in a stepwise manner as $\kappa_{\bm R}/\beta$ increases. It is also observed that there are two points where the rate of decrease in $N^{\rm str}(\kappa_{\bm R}/\beta)$ is particularly significant. Furthermore, approximately 80\% of the structures are found within the range $0 \leq \kappa_{\bm R}/\beta \leq 0.5$, indicating that most structures have a gentle slope around the minimum of the free energy $\Psi(\bm R, \mu)$. The proportion of structures with $\kappa_{\bm R}/\beta < 0$, which are outside the range of Fig. \ref{6_by_6_changes_of_num_of_struct_with_chi}, is approximately 4\% and low. Additionally, there are no structures with $\kappa_{\bm R}/\beta \geq 9.5$, as the maximum value of $\kappa_{\bm R}/\beta$ is 9.3971.

\subsection{Changes in the secondary structures densities according to environmental robustness}
\label{subsec:changesintheprop}

Next, we illustrate the changes in the proportion of secondary structures, the ratio of the number of amino acid residues in each secondary structure to the total number of residues, $N$ within groups of lattice protein structures as the stability $\kappa_{\bm R}$ increases incrementally. We also show the results for the quantity $\kappa_{\bm R}$ divided by $\beta$. 

\begin{figure*}[tb]
\begin{center}
\vspace{5mm}
\includegraphics[width=1.0\linewidth]{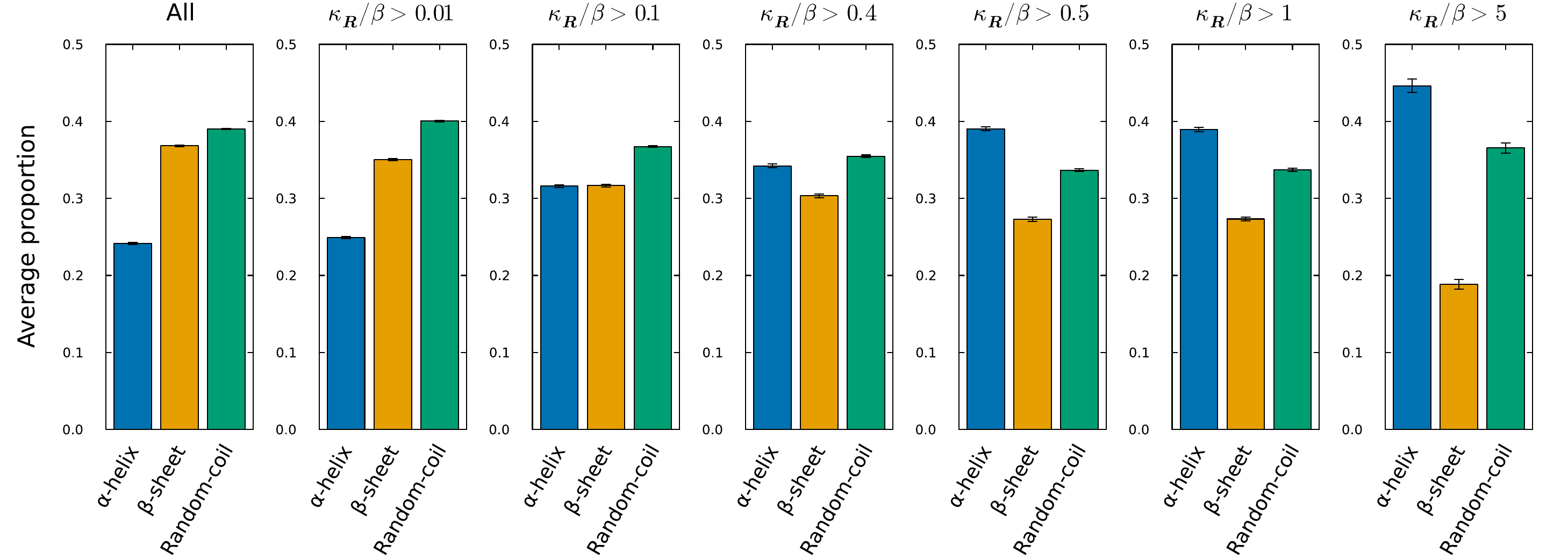}
\end{center}
\caption{The changes in $\alpha$-helix proportion, $\beta$-sheet proportion, and random-coil proportion in sets of lattice protein structures with $\kappa_{\bm R}/\beta$ values greater than specified thresholds for a $6 \times 6$ lattice protein. Each bar graph represents the average value of each secondary structure proportion for all structures exceeding the given $\kappa_{\bm R}/\beta$ threshold, with error bars represent the standard error of the mean for each region. The plot shows the secondary structure proportions across all $6 \times 6$ structure patterns, with subsequent values shown for $\kappa_{\bm R}/\beta$ thresholds incrementally changed to 0.01, 0.1, 0.4, 0.5, 1, and 5. The sample sizes for each group (i.e., the number of lattice protein structures satisfying each inequality) are, in order of increasing lower bound of $\kappa_{\bm R}/\beta$—that is, from the second graph from the left in Fig. 4—18767, 10460, 5492, 4080, 4048, and 460. Each bar graph uses blue for $\alpha$-helix proportion, yellow for $\beta$-sheet proportion, and green for random-coil proportion. Since any non-$\alpha$-helices and non-$\beta$-sheets structures are categorized as random-coils, the sum of the three colored bar graphs totals 1.}
\label{secondary_structure_change}
\end{figure*}

For each increment in $\kappa_{\bm R}/\beta$, we plot the average values of $\alpha$-helix proportion, $\beta$-sheet proportion, and random-coil proportion within the set of structures that have a $\kappa_{\bm R}/\beta$ greater than the current $\kappa_{\bm R}/\beta$ value. Fig.\ref{secondary_structure_change} shows its distributions of secondary structures. We also set the inverse temperature at $\beta = 10$.

Looking at Fig.\ref{secondary_structure_change}, as $\kappa_{\bm R}/\beta$ increases, $\alpha$-helix proportion increases while $\beta$-sheet proportion decreases. Random-coil proportion remain neutral to $\kappa_{\bm R}/\beta$ changes. This result indicates that structures with a higher $\alpha$-helix proportion tend to have higher $\kappa_{\bm R}/\beta$ values and those with a higher $\beta$-sheet proportion tend to have lower $\kappa_{\bm R}/\beta$ values. Therefore, it can be concluded that proteins with a higher $\alpha$-helix proportion are more robust over generations in response to perturbations in environmental conditions.

\subsection{Empirical validation using real protein structure database}
\label{subsec:EmpiricalValidation}
We tested our the theoretical prediction shown in\ref{subsec:changesintheprop} using publicly available protein databases. The validation method involved defining a group of proteins that, according to our research, exhibit high structural robustness against environmental perturbations across multiple generations. We then compared the secondary structure content of these proteins to that of a randomly selected group of proteins.

Although this validation method is relatively straightforward, determining which proteins should be categorized as possessing high evolutionary robustness remains a challenging issue. Ideally, such criteria would be quantitatively specified; however, this is extremely difficult in practice. In our study, we define a set of proteins generally regarded as evolutionarily robust. We select such proteins based on the following conditions: (i) they are found in all three domains of life (archaea, bacteria, and eukaryotes) and (ii) they perform a fundamental biological function. Proteins that meet these conditions have likely maintained their function since the very early stages of life. Given the strong correlation between protein structure and function, it is reasonable to hypothesize that such proteins possess high structural robustness over numerous generations in response to environmental perturbations. In this study, we selected the following eight protein types as examples of the Robust proteins: (1) Ribosomal proteins, (2) DNA polymerases, (3) DNA helicases, (4) Ribonucleotide reductases , (5) RNA polymerases, (6) Transcription factors, (7) Aminoacyl-tRNA synthetases, and (8) ATP synthases. These eight proteins are present in all three domains of life and are involved in the replication, preservation, transcription, and translation of genetic information or in fundamental metabolic processes. Hereafter, we call these eight proteins as ``Robust proteins." As a control group, we randomly select proteins from the PDB. We call these proteins ``Random proteins."

We here briefly explain how the PDB files for “Robust proteins” and “Random proteins” were selected (for more detailed information, see Appendix \ref{sec:appendixB}).
For Robust proteins, each protein name was used as a keyword in a PDB text search. From the list of hits, the top 10,000 entries were taken. Out of these 10,000 entries, only those whose titles contained the protein name were selected.
For Random proteins, the procedure was as follows. First, we carried out the PDB text search with the keyword “Proteins,” and the PDB IDs of all resulting entries were downloaded. When searching for “Proteins,” nucleic acid and peptides are excluded, meaning only protein molecules are returned. Due to computational constraints in the subsequent secondary-structure analyses, we randomly selected 20,000 PDB IDs from all downloaded PDB IDs.
In both the Robust proteins and Random proteins selections, this method inevitably leaves a bias reflecting how many structures of each protein type are registered in the PDB. However, we ignore such a bias because that bias is the same for both sets.

Furthermore, we apply a redundancy reduction process to both Robust and Random proteins. Redundancy reduction was performed by clustering based on sequence similarity and selecting only the representative structures from each cluster. We used for the clustering CD-HIT (Cluster Database at High Identity with Tolerance) \cite{li2006cd}. CD-HIT is a tool widely used in protein science and the life sciences that clusters highly similar sequences and extracts representative sequences, thereby reducing redundancy in the dataset. For the Robust proteins, we carried out such redundancy reduction via CD-HIT for the PDB data set of each protein type after selection by title. We carried out the redundancy reduction for the randomly selected 20,000 PDB data for the Random proteins.

We then identified the secondary structures formed by individual amino acids using the DSSP (Dictionary of Secondary Structure in Proteins) algorithm \cite{touw2015series, kabsch1983dictionary}, which assigns secondary structure to the amino acids using the backbone hydrogen bonds. DSSP classifies amino acids into eight secondary structure types, each represented by a single-letter code \cite{kabsch1983dictionary}. Our analysis considered only ``H" as an $\alpha$-helix, only ``E" as a $\beta$-sheet, and all other characters as a random-coil. This is because conservative choice like this is more appropriate when considering the correspondence with lattice proteins (see Appendix \ref{sec:appendixB} for a detailed discussion). Structures with missing residue coordinate information in the PDB could not be processed by DSSP and were thus excluded.
Furthermore, we exclude the structures that DSSP reported 0\% for both $\alpha$-helix and $\beta$-sheet proportions. The main reason for this step is that the present theory focuses its analysis on compact proteins that possess secondary structures such as $\alpha$-helices and $\beta$-sheets. After these processing steps, the final number of proteins used for computing the comparative distribution of secondary structure proportions was 557 for the Robust proteins and 4,987 for the Random proteins. Details on data selection from the PDB, preprocessing procedures, and DSSP analysis are summarized in Appendix 
\ref{sec:appendixB}.

\begin{figure*}[tb]
\begin{center}
\vspace{5mm}
\includegraphics[width=1.0\linewidth]{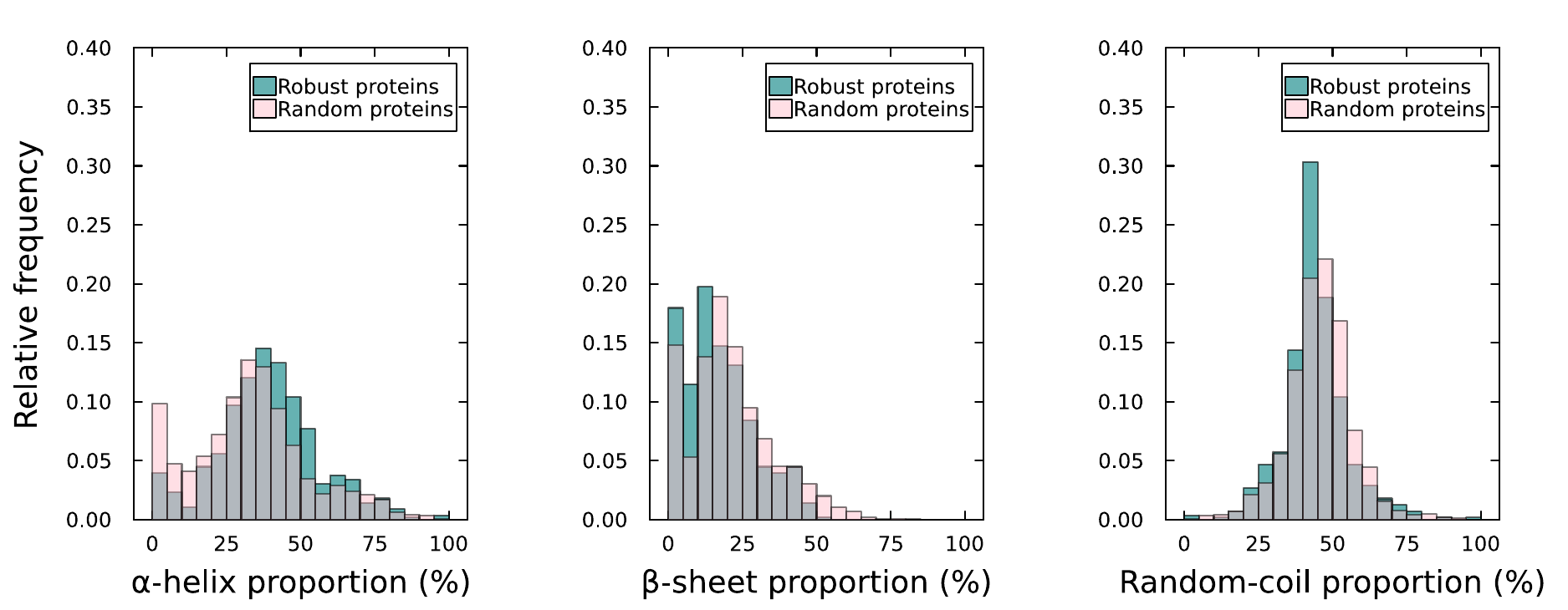}
\end{center}
\caption{The histograms compare the $\alpha$-helix proportion, $\beta$-sheet proportion, and random-coil proportion between  the Robust proteins (teal) and the Random proteins (pink). The vertical axes of the three histograms represent the relative frequency of each secondary structure proportions of proteins of both Robust and Random group. In all figures, the bin width is set to 5\%. The sample sizes for each secondary structure category are 557 for the Robust proteins and 4,987 for the Random proteins. At first glance, there appear to be no major differences between the two protein groups across all three secondary structures. However, for $\alpha$-helix proportion and $\beta$-sheet proportion, the histograms for the Robust proteins are systematically shifted to the right and left, respectively. This observation supports our theoretical predictions. In the case of random-coil proportion, the difference is smaller than that observed for $\alpha$-helix proportion and $\beta$-sheet proportion, but the histogram for the Robust proteins is slightly shifted to the left. Furthermore, for all three secondary structures, the overall shapes of the histograms for the Robust proteins and the Random proteins are similar. Notably, both datasets exhibit approximately Gaussian-shaped histograms in the case of random-coil proportion.}
\label{real_data_dist}
\end{figure*}

We draw the results in Fig. \ref{real_data_dist}. Fig.  \ref{real_data_dist} shows overlaid histograms of secondary structure proportions for Robust and Random proteins.
As shown in Fig. \ref{real_data_dist}, there are no substantial differences between the Robust proteins and the Random proteins with respect to the overall proportions of $\alpha$-helix, $\beta$-sheet, and random-coil. However, there are slight differences in all three distributions. Specifically, the distribution of $\alpha$-helix proportion for the Robust proteins is shifted slightly to the right, whereas the distribution of $\beta$-sheet proportion is correspondingly shifted slightly to the left. These observations support our theoretical prediction (Fig. \ref{secondary_structure_change}). Table \ref{tab:EmpiricalValidationTable} provides the mean and standard deviation (STD) of the secondary structure proportions for each group. Table \ref{tab:EmpiricalValidationTable} Table \ref{tab:EmpiricalValidationTable} qualitatively supports our theoretical prediction in \ref{subsec:changesintheprop}.

However, the information presented above alone is insufficient to determine whether there are statistically significant differences in the densities of each secondary structure.
Therefore, we turn to statistical hypothesis testing.
As shown in Fig.~\ref{real_data_dist}, the $\alpha$-helix proportion and $\beta$-sheet proportion data do not follow any clear parametric distribution.
Consequently, we employed a hypothesis test based on the bootstrap method 
\cite{efron1994introduction} (hereafter referred to as the bootstrap test).

The bootstrap test is a statistical approach rooted in bootstrapping, in which multiple samples are generated by resampling (with replacement) from a single original dataset to estimate the sampling distribution.
This non-parametric method allows testing even when the functional form of the population distribution is unknown.
In the bootstrap test, the null hypothesis $\mathrm{H}_0$ posits that there is no difference between the mean values of the two groups.
To evaluate $\mathrm{H}_0$, we begin by merging the datasets for Robust and Random proteins into a single dataset, regarding it as a surrogate for the population distribution under $\mathrm{H}_0$.
We then draw bootstrap samples by resampling with replacement from this merged dataset, selecting the same number of samples as in the original two datasets.
This process simulates data acquisition under $\mathrm{H}_0$ and is repeated many times (100,000 in our case) to generate a histogram of the mean differences.
If the mean difference observed in the original two groups lies within a tail of this histogram whose frequency is below the significance level, we reject $\mathrm{H}_0$.
This constitutes the procedure for our hypothesis test.

Details of the methodology and results of the bootstrap test can be found in Appendix \ref{sec:appendixC}.
For the random-coil proportion data, we applied the t-test since its 
shape of histogram closely resembled a Gaussian distribution. 
The significance level was set to 0.01 for the statistical hypothesis tests in all cases of $\alpha$-helix proportion, $\beta$-sheet proportion, and random-coil proportion.

The p-values evaluated by the statistical tests 
were 0.0, 0.0, and 0.0002 for $\alpha$-helix, $\beta$-sheet, 
and random-coil proportions, respectively.
Thus, the results indicate, with significance level 0.01,  
that there are
differences in mean values in all the three cases.
These results support our theoretical predictions for the $\alpha$-helix proportion and $\beta$-sheet proportion data. As for the random-coil proportion, the theoretical prediction shown in Fig. \ref{secondary_structure_change} suggests a slightly decreasing trend in the range of $1<\kappa_{\bm R}/\beta$, which is consistent by the empirical data. However, in the region where $5<\kappa_{\bm R}/\beta$—represented by the rightmost bar in Fig. \ref{secondary_structure_change}—the random-coil proportion increases, which contradicts the empirical results. This discrepancy is likely due to structural differences between random coils in lattice proteins and those defined in real proteins in this study. DSSP classifies each amino acid residue into eight structural types. In our analysis, we categorized certain structural elements as random coil while excluding isolated $\beta$-bridges (residues forming one-to-one hydrogen bonds with adjacent residues but not contributing to a sufficiently extended $\beta$-sheet) and hydrogen-bond-mediated turn structures from the $\beta$-sheet category. In contrast, 2D lattice proteins lack the resolution to accurately define or represent such structures. Additionally, 2D lattice proteins generally contain more $\alpha$-helix-like and $\beta$-sheet-like structures than real proteins. This tendency may have led to overestimating the contribution of $\alpha$-helix and $\beta$-sheet content to the free energy change $\Psi(\bm R, \mu)$. Therefore, the discrepancy in random-coil results should not necessarily be interpreted as a fundamental flaw in our proposed evolutionary theory.

\begin{table}[htb]
  \begin{center}
  \caption{The mean and standard deviation (STD) values of the $\alpha$-helix, $\beta$-sheet, and random-coil proportions for both the Robust proteins and the Random proteins.}
	\begin{tabular}{ccccc} \hline \hline
	  Secondary structure & Robust or Random& Mean (\%) \hspace{5mm}& STD      \\\hline
	\multirow{2}{*}{$\alpha$-helix} & Robust & 39.18 & $\pm 17.35$\\
	& Random & 33.16 & $\pm 19.59$ \\ \hline
	\multirow{2}{*}{$\beta$-sheet} &Robust & 16.79 & $\pm 12.17$\\
	&Random & 20.98 & $\pm 14.22$\\ \hline
	\multirow{2}{*}{Random-coil} &Robust & 44.03 & $\pm 10.73$\\
	&Random & 45.86 & $\pm 10.82$\\ \hline \hline
	\end{tabular}
	\label{tab:EmpiricalValidationTable}
  \end{center}
\end{table}

\section{DISCUSSION}
\label{sec:discussion}

First, we note some overall considerations regarding our statistical mechanics theory framed within Bayesian learning for evolution. Our proposed evolutionary theory does not optimize phenotypes to given environmental conditions. That is because the stability analysis around the minimum of the free energy $\Psi(\bm R, \mu)$ pertains to the stability around the minimum for environmental conditions $\mu$ given a structure (phenotype) $\bm R$ and not the reverse. According to the neutral theory of molecular evolution\cite{kimura1968evolutionary}, the evolution of organisms (molecular evolution) is not necessarily a product of optimization under given environmental conditions. Our theory does not contradict these important evolutionary concepts.

The evolutionary significance of $\kappa_{\bm R}$ is, directly, the stability to change in the amino acid sequence $\bm \sigma$ (genotype) that forms the structure (phenotype) $\bm R$ over subsequent generations. Thus, it cannot necessarily be said to explain the robustness to change of a structure over many generations. Furthermore, $\kappa_{\bm R}$ pertains explicitly to the stability against perturbations in environmental conditions $\mu$, and does not directly inform about the stability against significant changes in $\mu$. Therefore, while $\kappa_{\bm R}$ may be better suited to explaining incremental microevolution, it may not be as applicable to macroevolution. However, it is not uncommon for macroevolution in organisms to be understood as an accumulation of microevolutions; our theory and analysis of $\kappa_{\bm R}$ could still play a significant role in understanding large evolutionary changes.

$\kappa_{\bm R}$ can also be interpreted as part of a broader concept of robustness that includes environmental robustness, especially considering the functionality of gene regulatory networks. In this sense, proposing $\kappa_{\bm R}$ in an evolutionary context is meaningful. However, it should be noted, as emphasized in Chapter II, that all the measures proposed in this study, such as $\chi_{\bm R}$ and $\kappa_{\bm R}$, averages over the amino acid sequences $\bm \sigma$. Thus, the effects of mutational robustness across successive generations on the structure $\bm R$ remain unclear. This study exclusively analyzed the effects of environmental fluctuations on evolution without integrating comprehensive evolutionary theories that include mutations, natural selection, or genetic drift, which are changes in genes unrelated to natural selection. To include theories that account for mutations and genetic drift, the former would require discussing the stability of quantities that retain amino acid sequence pattern dependencies without integrating over amino acid sequences for proteins. For the latter, it would be necessary to consider quantities or models where hydrophobicity fluctuates solely due to random effects while being constant to changes in the environmental condition $\mu$. Constructing such theories remains a task for future research.

The result of Fig.\ref{6_by_6_changes_of_num_of_struct_with_chi} indicates that increasing of $\kappa_{\bm R}$ leads to a very rapid reduction in the phenotypic space. It is unclear what proportion of the total structural space, including evolutionary non-viable protein structure patterns, is occupied by actual protein structures, so whether the results from Fig.\ref{6_by_6_changes_of_num_of_struct_with_chi} accurately explain the dimensional reduction of phenotypes from the perspective of natural selection remains uncertain. However, the fact that evolved phenotypes constitute only a portion of the entire possible phenotypic space can indeed be explained by our proposed metric of robustness to change in phenotypes over subsequent generations in response to perturbations in environmental conditions $\mu$.

Similarly, Fig.\ref{secondary_structure_change} lacks direct biological evidence. However, a conventional study elucidated that $\alpha$-helices exhibit higher mutational robustness than $\beta$-sheets \cite{abrusan2016alpha}. Of course, the relationship between mutational robustness and our results is unclear. However, the fact that $\alpha$-helices and $\beta$-sheets behave differently regarding structural mutational robustness, which is evolutionarily significant, is likely essential. Given the observed variation in the proportions of $\alpha$-helices and $\beta$-sheets among proteins, it is plausible to suggest that these differences may be associated with differences in protein function. Considering that protein function closely relates to evolvability, the differing dependencies on the proportions of $\alpha$-helices and $\beta$-sheets suggest that our proposed stability measure, $\kappa_{\bm R}$, is likely to be evolutionarily meaningful.

The empirical validation discussed in \ref{subsec:EmpiricalValidation} shows that, while statistical hypothesis testing indicates that the $\alpha$-helix proportion and $\beta$-sheet proportion distributions of Robust proteins and Random proteins are significantly different, the observed differences are not particularly large at first glance. This may suggest that existing proteins, having endured long evolutionary histories, are inherently structurally robust. In other words, as inferred from Fig. \ref{6_by_6_changes_of_num_of_struct_with_chi}, the structures of evolved proteins occupy an exponentially small subset of the physically possible structural space. Under this assumption, selecting proteins that are presumably evolutionarily robust from biological databases such as the PDB, as done in this study, may inherently make it difficult to identify a clear distinction from random samples.

\section{CONCLUSION}
Here, we give the conclusion of our study. We proposed an evolutionary generation model for protein structures by statistical mechanics based on Bayesian learning framework. We considered the chemical potential surrounding proteins as an environmental condition. We discussed the stability of the free energy as a function of protein structure and environmental conditions, as defined in Eq. (\ref{evolutionary free energy}). This stability refers to the robustness to change in hydrophobicity (the proportion of hydrophobic amino acids, analogous to magnetization in magnetic materials), determined by the amino acid sequences that design (fold into) a specific protein structure with high probability. Since changes in hydrophobicity can lead to structural changes, this stability can be considered as the stability of a given protein structure against environmental perturbations over subsequent generations. Consequently, in a 2D square lattice protein model composed of 36 residues, we found that structures with a certain level of stability in their free energy are very rare in the entire structural space, with a higher proportion of $\alpha$-helices and a lower proportion of $\beta$-sheets. Furthermore, comparing secondary structure densities between the Robust proteins (proteins that are likely to be evolutionarily robust from a biological perspective.) and the Random proteins (randomly selected structures from PDB) based on empirical data qualitatively supports the theoretical predictions for $\alpha$-helix and $\beta$-sheet structures.

\section*{ACKNOWLEDGMENTS}
This work was supported by JSPS KAKENHI Grant Number 23K19996 (T. T.). The authors are grateful to Macoto Kikuchi, Ayaka Sakata, and Takashi Takahashi for illuminating discussions and helpful comments. This work was also supported by JSPS KAKENHI Grant Number 22H00406 (G. C.), and JSPS KAKENHI Grant Number 22H05117 and JST CREST JPMJCR1912 (Y. K.)

\appendix
\section{The calculations of hydrophobicity and susceptibility using Belief propagation}
\label{sec:appendixA}
\renewcommand{\theequation}{A\arabic{equation}}
\setcounter{equation}{0}


\subsection{The calculations of $\langle \sigma_{i}\rangle_{\textbf{\textit{R}}}$ and $\langle \sigma_{i}\rangle$}

In order to calculate the hydrophobicity of single structure, one needs to obtain the posterior average of single residue $\left<\sigma_{i}\right>_{\bm R}$ given by
\begin{equation}
\label{h_R single residue}
    \left<\sigma_{i}\right>_{\bm R} = \sum_{\bm \sigma} \sigma_{i} \hspace{1mm}p({\bm \sigma}|\bm R, \mu).
\end{equation}
If the posterior $p({\bm \sigma}|\bm R, \mu)$ is decoupled to each residue expressed as follows by using the set of residues without $\sigma_{i}$, $\bm \sigma \backslash i$
\begin{align}
\label{marginal posterior}
    p_{i}({\sigma_{i}}|\bm R, \mu) = \sum_{\bm \sigma \backslash i} p({\bm \sigma}|\bm R, \mu),
\end{align}
then, one can rewrite Eq. (\ref{h_R single residue}) very simple form as follows:
\begin{align}
    \left<\sigma_{i}\right>_{\bm R} &= \sum_{\sigma_{i} = 0,1} \sigma_{i} \hspace{1mm}p_{i}({\sigma_{i}}|\bm R, \mu)\\
    & = p_{i}(\sigma_{i} = 1|\bm R, \mu).
\end{align}

Belief propagation (BP) can obtain the marginal distribution $p_{i}({\sigma_{i}}|\bm R, \mu)$ by using following update rules, 
\begin{align}
\label{BP update rule 1}
    \tilde{\nu}_{a \rightarrow i}^{(t)} (\sigma_{i}) = \frac{1}{Z_{a \rightarrow i}} \sum_{\sigma_{j(i)}} e^{\beta \sigma_{i}\sigma_{j(i)}} \nu_{j(i) \rightarrow a}^{(t)} (\sigma_{j(i)}), \\
    \label{BP update rule 2}
\nu_{i \rightarrow a}^{(t + 1)} (\sigma_{i}) = \frac{1}{Z_{i \rightarrow a}} e^{\beta\mu(1 - \sigma_{i})} \prod_{b \in \partial_{i} \setminus a} \tilde{\nu}_{b \rightarrow i}^{(t)} (\sigma_{i}).
\end{align}
In Eqs. (\ref{BP update rule 1}) and (\ref{BP update rule 2}), the \textit{beliefs} or \textit{messages} $\tilde{\nu}_{a \rightarrow i}^{(t)} (\sigma_{i})$ and $\nu_{i \rightarrow a}^{(t + 1)} (\sigma_{i})$ are the probability from the $a$-th contact to the $i$-th residue and the probability from the $i$-th residue to the $a$-th contact, respectively. The subscripts $a, b, \cdots$ are indices on contacts, and the upper right subscript is the number of steps in the BP algorithm. The symbol $\partial_{i}$ denotes the index set of contacts related to residue $\sigma_{i}$. The constants $Z_{a \rightarrow i}$ and $Z_{i \rightarrow a}$ are the normalizing constants of each distribution function. The residue index $j(i)$ denotes the index that contacts with $i$-th residue. In the lattice HP model, all residue-residue interactions are two-body. Thus, this index $j(i)$ is unique to $i$.

The derivation of the BP update rules (\ref{BP update rule 1}) and (\ref{BP update rule 2}) are somewhat technical, so they are not presented here. Please refer to the appendix in our previous work\cite{takahashi2022cavity} for information of the derivation of the above BP update rules.

If one properly defines $\nu_{i \rightarrow a}^{(t = 0)}(\sigma_{i})$ as the initial condition (in this study, we use 0.5.) and computes Eqs. (\ref{BP update rule 1}) and (\ref{BP update rule 2}) at each step for all combinations $(i, a)$, after sufficient iterations $t_{\rm max}$, the following belief:
\begin{eqnarray}
\label{marginal posterior by BP message}
	\nu_{i}(\sigma_{i}) = \frac{1}{Z_{i}}\prod_{a \in \partial_{i}} \tilde{\nu}_{a \rightarrow i}^{(t_{\rm max})} (\sigma_{i}),
\end{eqnarray}
converges to the marginal distribution $p_{i}(\sigma_{i} | \bm R,\mu)$. In Eq. (\ref{marginal posterior by BP message}) where $Z_{i}$ is the normalization constant.

In order to obtain another hydrophobicity $h$, one needs to calculate the joint average of $\sigma_{i}$. It is expressed as follows.
\begin{align}
    \left<\sigma_{i}\right> &= \sum_{\bm R}\sum_{\bm \sigma} \sigma_{i} \hspace{1mm}p(\bm R, \bm \sigma | \mu)\\
    & = \sum_{\bm R}\sum_{\bm \sigma} \sigma_{i} \hspace{1mm}p(\bm R | \mu)p(\bm \sigma | \bm R, \mu)\\
    & = \sum_{\bm R}\sum_{\bm \sigma} \sigma_{i} \hspace{1mm}p(\bm \sigma | \bm R, \mu)p(\bm R | \mu)\\
    \label{A11}
    & = \sum_{\bm R} \left<\sigma_{i}\right>_{\bm R} p(\bm R | \mu)\\
     \label{A12}
    & = \frac{\sum_{\bm R} \left<\sigma_{i}\right>_{\bm R} Y(\bm R;\beta,\mu)}{\sum_{\bm R}Y(\bm R;\beta,\mu)}.
\end{align}
In Eq. (\ref{A11}) to Eq. (\ref{A12}), we used Eq. (\ref{marginal likelihood by sequence partition function}). Therefore, one has to calculate the sequence partition function $Y(\bm R;\beta,\mu) = \sum_{\bm \sigma}e^{-\beta H(\bm R, \bm \sigma; \mu)}$ and the posterior average of each residue $\left<\sigma_{i}\right>_{\bm R}$ for all lattice structure patterns. We have the all patterns of the self avoiding walks on 2D $N = 6\times6$ square. Thus, in this study, we can obtain the exact value of $\left< \sigma_{i}\right>$.

For more large size of lattice proteins, one has to carry out the efficient multi-canonical Monte Carlo methods suitable for exploring lattice protein structures\cite{chikenji1999multi, shirai2013multicanonical}. For the realistic protein structures, such a structural search is extremely difficult even with the use of the database\cite{berman2000protein}, because the structural search space has to involve the random structural patterns including structures that have not evolved.

By using BP, The sequence partition function $Y(\bm R;\beta,\mu) = \sum_{\bm \sigma}e^{-\beta H(\bm R, \bm \sigma; \mu)}$ is obtained from the Bethe free entropy $F_{B}(\tilde{\bm \nu}^{*}) = \log Y(\bm R;\beta,\mu)$ where $\tilde{\bm \nu}^{*}$ is the set of the contact to residue messages (Eq. (\ref{BP update rule 1})) after sufficient time steps. Free entropy is -$\beta$ times free energy.

The Bethe free entropy $F_{B}(\tilde{\bm \nu}^{*})$ is the free entropy under the Bethe-approximation. In the lattice HP model using the Hamiltonian (1), $F_{B}(\tilde{\bm \nu}^{*})$ is given by
\begin{align}
\label{Bethe free energy}
    F_{B}(\tilde{\bm \nu}^{*}) &= \sum_{a=1}^{M} \log Z_{a} + \sum_{i=1}^{N} \log Z_{i} - \sum_{ia} \log Z_{ia},\\
    \label{Za}
    Z_{a} &:= \sum_{\sigma_{i}\sigma_{j(i)}} e^{\beta \sigma_{i}\sigma_{j(i)}} \prod_{i \in \partial a} \nu_{i \rightarrow a}^{*}(\sigma_{i}),\\
    \label{Zi}
    Z_{i} &:= \sum_{\sigma_{i}} e^{\beta \mu (1-\sigma_{i})} \prod_{b \in \partial i} \tilde{\nu}_{b \rightarrow i}^{*}(\sigma_{i}),\\
    \label{Zia}
    Z_{ia} &:= \sum_{\sigma_{i}} \tilde{\nu}_{a \rightarrow i}^{*}(\sigma_{i}) \nu_{i \rightarrow a}^{*}(\sigma_{i}),
\end{align}
where $ia$ denotes index of elements of the set in which all combinations of residues and contacts, and the symbol $\partial a$ denotes the index set of residues related to contact $a$. The symbol $M$ denotes the number of contacts. The messages $\nu_{i \rightarrow a}^{*}(\sigma_{i})$ and $\nu_{a \rightarrow i}^{*}(\sigma_{i})$ are the converged $i$-th residue to $a$-th contact message and the converged $a$-th contact to $i$-th residue message, respectively. The derivation of Eqs. (\ref{Bethe free energy}) \textasciitilde (\ref{Zia}) for general cases involves a technical and lengthy explanation, which is beyond the scope of this paper. For further details, please refer to the representative texts\cite{mezard2001bethe, mezard2009information}. Then, we obtain
\begin{align}
\label{sequence partition function by BP}
    Y(\bm R;\beta,\mu) = \frac{\left( \prod_{a} Z_{a} \right) \left( \prod_{i} Z_{i} \right)} {\prod_{ia} Z_{ia}}.
\end{align}

In order to obtain $\mu_{EB}$ for each structure, we derive the minimization condition of free energy of a structure $\Psi(\bm R, \mu)$ defined by Eq. (\ref{evolutionary free energy}). The minimization condition with respect to $\mu$ : $(\partial / \partial \mu) \Psi(\bm R, \mu) = 0$ becomes as follows:
\begin{align}
\label{minimization condition}
\left< \sum_{i=1}^{N}\sigma_{i} \right>_{\bm R} = \left< \sum_{i=1}^{N}\sigma_{i} \right>.
\end{align}
The minimization parameter $\mu_{EB}$ satisfies Eq. (\ref{minimization condition}). Thus, one can get $\mu_{EB}$ by computing $\sum_{i} \left< \sigma_{i} \right>_{\bm R}$ and $\sum_{i} \left< \sigma_{i} \right>$ through the method explained above for each structure. We used the bisection method to compute left hand side and right hand side of Eq. (\ref{minimization condition}).

\subsection{The calculations of $\chi_{\textbf{\textit{R}}}$ and $\chi$}

In order to obtain the susceptibility $\chi_{\bm R}$, one has to calculate the residue correlation function $\left< \sigma_{i} \sigma_{j}\right>_{\bm R}$ given by
\begin{align}
    \left<\sigma_{i}\sigma_{j}\right>_{\bm R} = \sum_{\bm \sigma} \sigma_{i}\sigma_{j} \hspace{1mm}p({\bm \sigma}|\bm R, \mu).
\end{align}
If $\sigma_{i}$ and $\sigma_{j}$ are not statistically independent, i.e., when there is at least one path connecting $\sigma_{i}$ and $\sigma_{j}$ on the contact graph, one has to think the joint probability
\begin{align}
    p_{i,j}(\sigma_{i}, \sigma_{j} | \bm R, \mu) = p_{i | j}(\sigma_{i} | \sigma_{j}, \bm R, \mu) p_{j}(\sigma_{j} | \bm R, \mu),
\end{align}
where the conditional probability, $p_{i | j}(\sigma_{i} | \sigma_{j}, \bm R, \mu)$ is given by 
\begin{align}
    p_{i | j}(\sigma_{i} | \sigma_{j}, \bm R, \mu) = \frac{p_{i,j}(\sigma_{i}, \sigma_{j} | \bm R, \mu)}{p_{j}(\sigma_{j} | \bm R, \mu)}.
\end{align}
If the following marginalization can be carried out
\begin{align}
    p_{i | j}(\sigma_{i} | \sigma_{j}, \bm R, \mu) = \sum_{\bm \sigma \backslash i}p(\bm \sigma | \sigma_{j} ,\bm R,\mu),
\end{align}
one can then obtain $\left< \sigma_{i} \sigma_{i}\right>_{\bm R}$ following quite simple form
\begin{align}
    \left< \sigma_{i} \sigma_{i}\right>_{\bm R} &= \sum_{\sigma_{i}}\sum_{\sigma_{j}} \sigma_{i}\sigma_{j} p_{i | j}(\sigma_{i} | \sigma_{j}, \bm R, \mu) p_{j}(\sigma_{j} | \bm R, \mu)\\
    &= p_{i | j}(\sigma_{i} = 1 | \sigma_{j}=1, \bm R, \mu) p_{j}(\sigma_{j}=1 | \bm R, \mu).
\end{align}
BP is also able to calculate the conditional marginal $p_{i | j}(\sigma_{i} | \sigma_{j}, \bm R, \mu)$. The procedure is as follows: one uses the conditional messages $\tilde{\nu}_{a \rightarrow i|j}^{(t)} (\sigma_{i} | \sigma_{j} = \sigma)$ and $\nu_{i|j \rightarrow a}^{(t + 1)} (\sigma_{i} | \sigma_{j} = \sigma)$ instead of normal BP messages Eqs.(\ref{BP update rule 1}) and (\ref{BP update rule 2}), where $\sigma$ is the realization of $\sigma_{j}$. Thus, in the current case $\sigma = 1$, the normal BP messages change to following conditional forms
\begin{align}
\label{conditional BP messages 1}
     \tilde{\nu}_{a \rightarrow i|j}^{(t)} (\sigma_{i} | \sigma_{j} = 1)&= \frac{1}{Z_{a \rightarrow i|j}} \sum_{\sigma_{k(i)}} e^{\beta \sigma_{i}\sigma_{k(i)}} \nu_{k(i)|j \rightarrow a}^{(t)} (\sigma_{k(i)}|\sigma_{j} = 1),\\
     \label{conditional BP messages 2}
     \nu_{i|j \rightarrow a}^{(t + 1)} (\sigma_{i} | \sigma_{j} = 1)&= \frac{1}{Z_{i|j \rightarrow a}} e^{\beta\mu(1 - \sigma_{i})} \prod_{b \in \partial_{i} \setminus a} \tilde{\nu}_{b |j\rightarrow i}^{(t)} (\sigma_{i} | \sigma_{j} = 1). 
\end{align}
In Eq. (\ref{conditional BP messages 1}), we use $k(i)$ as the index of residue that contact with $\sigma_{i}$ to avoid confusion with $j$ of current context. The symbol $Z_{a \rightarrow i|j}$ and $Z_{i|j \rightarrow a}$ are the  normalization constants for the corresponding messages. As with the normal BP described earlier, the following belief:
\begin{eqnarray}
\label{A26}
    \nu_{i|j}(\sigma_{i} | \sigma_{j} = 1) = \frac{1}{Z_{i|j}}\prod_{a \in \partial_{i}} \tilde{\nu}_{a \rightarrow i|j}^{(t_{\rm max})} (\sigma_{i} | \sigma_{j} = 1),
\end{eqnarray}
converges to the conditional marginal $p_{i | j}(\sigma_{i} | \sigma_{j}=1, \bm R, \mu)$. The symbol $Z_{i|j}$ in Eq. (\ref{A26}) denotes the normalization constant of this message.

One can calculate $\chi$ by using following formula
\begin{align}
\label{A27}
    \left< \sigma_{i} \sigma_{j} \right> = \frac{\sum_{\bm R} \left<\sigma_{i}\sigma_{j}\right>_{\bm R} Y(\bm R;\beta,\mu)}{\sum_{\bm R}Y(\bm R;\beta,\mu)}.
\end{align}
Eq. (\ref{A27}) is obtained by the same manner as the derivation process of Eq.(\ref{A12}).

\section{The details of the empirical validation using PDB}
\label{sec:appendixB}
Here, we summarize the details of the empirical data analysis presented in \ref{subsec:EmpiricalValidation}. First, we describe the selection process of PDB data for all proteins used in the analysis. For Robust proteins, we obtained structural data to generate the histograms in Fig. \ref{real_data_dist} and Table \ref{tab:EmpiricalValidationTable} following these steps for each of the eight protein categories: (1) Ribosomal proteins, (2) DNA polymerases, (3) DNA helicases, (4) Ribonucleotide reductase, (5) RNA polymerases, (6) Transcription factors, (7) Aminoacyl-tRNA synthetases, and (8) ATP synthases. The procedure was as follows:
\begin{enumerate}
    \item Perform a text search in the PDB.
    \item Retrieve the top 10,000 entries based on the search results.
    \item Select entries containing the specific protein name from the remaining dataset.
    \item Perform clustering based on sequence similarity with a threshold of 50\% and extract only the representative structures from each cluster.
\end{enumerate}
In Step 2, Ribosomal proteins and Aminoacyl-tRNA synthetases were not strictly excluded from the dataset based on the exact order of words in the PDB entry titles. Specifically, for ribosomal proteins, entries containing phrases such as ``ribosomal subunit" (e.g., ``large ribosomal subunit") were included in the dataset for subsequent analysis.
For aminoacyl-tRNA synthetases, these proteins are typically registered in the format ``$\textit{specific amino acid}$ $+$ -tRNA synthetase" (e.g., Glutamyl-tRNA synthetase, Cysteinyl-tRNA synthetase, and Valyl-tRNA synthetase). Therefore, any entries containing the phrase ``$\textit{any word}+$ -tRNA synthetase" were included in the dataset for further analysis.

Additionally, during DSSP secondary structure analysis, some structures were excluded following these steps:
\begin{enumerate}
    \item Remove ligands and perform DSSP analysis on the representative structures.
    \item Exclude structures where both $\alpha$-helix density and $\beta$-sheet density were 0\%.
\end{enumerate}
Structures with missing residue coordinate information in the PDB could not be processed by DSSP and were thus excluded. Furthermore, we implemented the final exclusion step since this study focuses on proteins containing $\alpha$-helix and $\beta$-sheet structures. Upon manually inspecting PDB entries where DSSP reported 0\% for both $\alpha$-helix and $\beta$-sheet proportions, we found cases where helix or sheet structures were present. These instances were considered potential DSSP errors and were, therefore, not included in the final dataset. This selection process is summarized in the flowchart shown in Fig. \ref{fig:flowchart}. The final number of structures for each protein category is shown in Table \ref{tab:ProteinCounts}.
\begin{table}[h]
    \centering
    \begin{tabular}{l c}
        \hline \hline
        Protein category & Number of structures \\
        \hline
        Ribosomal protein & 88 \\
        DNA polymerase & 70 \\
        DNA helicase & 9 \\
        Ribonucleotide reductase & 22 \\
        RNA polymerase & 77 \\
        Transcription factor & 104 \\
        Aminoacyl-tRNA synthetase & 156 \\
        ATP synthase & 31 \\
        \hline
        Total & 557 \\
        \hline \hline
    \end{tabular}
    \caption{Number of structures for each protein category in the Robust proteins dataset.}
    \label{tab:ProteinCounts}
\end{table}

The selection process for PDB IDs in the Random proteins group was as follows:
\begin{enumerate}
    \item Perform a text search in the PDB using the keyword "Proteins" (to exclude nucleic acids and peptides).
    \item Randomly select 20,000 structures from the retrieved entries.
    \item Perform clustering based on sequence similarity with a threshold of 50\% and extract only the representative structures from each cluster.
\end{enumerate}
From the DSSP secondary structure analysis onward, the procedure was identical to that of the Robust proteins group. The final number of structures obtained was 4,987.

Finally, regarding DSSP analysis, DSSP classifies amino acids into the following eight secondary structure types, each represented by a single-letter code \cite{kabsch1983dictionary}:
\vspace{5mm}
\begin{itemize}
    \item H: $4_{13}$ helix (the common $\alpha$-helix)
    \item B: isolated $\beta$-bridge
    \item E: Extended $\beta$-strand (residues in parallel or antiparallel $\beta$-sheet)
    \item G: $3_{10}$ helix
    \item I: $\pi$-helix ($5_{16}$ helix)
    \item T: Hydrogen-bonded turn
    \item S: Bend
    \item <space> (often shown as a blank or ``.”): Loop or irregular/disordered structure not classified into any of the above categories
\end{itemize}
A detailed explanation of each classification can be found in \cite{kabsch1983dictionary}. The two helical structures other than the standard $\alpha$-helix differ in the number of amino acids per turn (i.e., the number of residues required for one complete helical turn). The isolated $\beta$-bridge is a single pairwise $\beta$-sheet-like interaction. The Bend is a pronounced change in the polypeptide backbone direction, defined by specific geometric criteria (e.g., a significant shift in the chain's trajectory). Unlike helices or sheets, it does not necessarily involve hydrogen bonding. Our analysis considered only ``H" as an $\alpha$-helix and only ``E" as a $\beta$-sheet. While a discussion could be made about including other classifications, our primary objective was to validate predictions from lattice protein models qualitatively. Given this purpose, we focus exclusively on the most definitive structures, ``H" and ``E," to ensure clarity and reliability in our comparison.

\begin{figure*}[htbp]
\centering
\begin{tikzpicture}[
  >=stealth
]

\tikzstyle{startstop} = [
  rectangle,
  rounded corners,
  minimum width=3cm,
  minimum height=1cm,
  text centered,
  draw=black
]
\tikzstyle{process} = [
  rectangle,
  minimum width=3cm,
  minimum height=1cm,
  text centered,
  draw=black
]
\tikzstyle{decision} = [
  diamond,
  draw=black,
  align=center,
  aspect=2.8,
  inner sep=1em
]

\node (start) [startstop] {Start};

\node (searchByText) [process, below of=start, yshift=-0.5cm]
  {Search on PDB by text. (eg: ``ribosomal protein")};

\node (get10000Title) [process, below of=searchByText, yshift=-0.5cm]
  {Get the titles of the top 10,000 hits.};

\node (elimByTitle) [decision, below of=get10000Title, yshift=-1.5cm]
  {Contains the search keyword?};

\node (cdHit) [process, below of=elimByTitle, yshift=-1.5cm]
  {Clustering by 50\% sequence similarity using CD-HIT};

\node (elimByClustering) [decision, below of=cdHit, yshift=-1.5cm]
  {Representative protein from each cluster?};

\node (dsspAnalysis) [process, below of=elimByClustering, yshift=-1.8cm]
  {Secondary structure calculation by DSSP};

\node (dsspSuccess) [decision, below of=dsspAnalysis, yshift=-1.3cm]
  {Success DSSP analysis?};

\node (alphaBetaCheck) [decision, below of=dsspSuccess, yshift=-2.8cm]
  {At least one of the $\alpha$-helix proportion or $\beta$-sheet proportion is non-zero.};

\node (end) [startstop, below of=alphaBetaCheck, yshift=-2.3cm]
  {End};

\node (noElimByTitle)        [process, right=2cm of elimByTitle]        {Eliminate};
\node (noElimByClustering)   [process, right=1.4cm of elimByClustering]   {Eliminate};
\node (noDsspSuccess)        [process, right=2.4cm of dsspSuccess]        {Eliminate};
\node (noAlphaBetaCheck)     [process, right=2.2cm of alphaBetaCheck]     {Eliminate};

\draw [->] (start) -- (searchByText);
\draw [->] (searchByText) -- (get10000Title);
\draw [->] (get10000Title) -- (elimByTitle);

\draw [->] (elimByTitle) -- node [right] {yes} (cdHit);
\draw [->] (cdHit) -- (elimByClustering);
\draw [->] (elimByClustering) -- node [right] {yes} (dsspAnalysis);
\draw [->] (dsspAnalysis) -- (dsspSuccess);
\draw [->] (dsspSuccess) -- node [right] {yes} (alphaBetaCheck);
\draw [->] (alphaBetaCheck) -- node [right] {yes} (end);


\draw [->](elimByTitle) -- node [above] {no} (noElimByTitle);    

\draw [->]
  (elimByClustering)
  -- node [above] {no}
     (noElimByClustering);

\draw [->]
  (dsspSuccess)
  -- node [above] {no}
     (noDsspSuccess);

\draw [->]
  (alphaBetaCheck)
  -- node [above] {no}
     (noAlphaBetaCheck);

\end{tikzpicture}
\caption{The selection flow of the PDB entries of the Robust proteins that we used for the empirical validation. From the sequence-similarity-based clustering onward, the same procedure also applies to the Random proteins.}
\label{fig:flowchart}
\end{figure*}

\section{Bootstrap Test}
\label{sec:appendixC}

Bootstrapping is a numerical procedure for evaluating the sampling distribution of a summary statistic (such as the sample mean, sample variance, or sample median) based on a single dataset \cite{efron1994introduction}.
This method relies on the idea that repeatedly resampling from the original dataset and computing the statistic each time provides a good estimate of the sampling distribution.
When performing statistical tests, we use this approach to construct the sampling distribution under the null hypothesis and then evaluate the 
$p$-value of the actually observed summary statistic.
This allows us to determine whether the null hypothesis should be accepted or rejected at a predetermined significance level.
A major advantage of this method is that it enables evaluation of the sampling distribution even when the underlying population distribution is unknown.
    
In this study, using the $\alpha$-helix proportion data and $\beta$-sheet proportion data shown in Fig.~\ref{real_data_dist}, we aim to demonstrate---at a given significance level---that the mean value for Robust proteins is higher in the former and lower in the latter.
In both datasets, the null hypothesis $\mathrm{H}_0$ states that there is no difference in the mean values of Robust and Random proteins, while the alternative hypothesis $\mathrm{H}_1$ posits that the mean value of Robust proteins is greater for $\alpha$-helix and smaller for $\beta$-sheet.
The sampling distribution under $\mathrm{H}_0$ can be constructed by bootstrapping as follows: 

\begin{enumerate}
    \item Merge the two datasets into a single dataset.
    \item Perform resampling with replacement from the merged dataset, drawing the same number of samples as in the original Robust proteins and Random proteins datasets, respectively. These bootstrap samples simulate the data acquisition process under the assumption that both Robust and Random proteins come from the same population distribution. Record the difference in mean values between these two bootstrap samples.
    \item Repeat Step 2 a total of 100,000 times to generate the distribution of the mean difference.
    \item Define the 
$p$-value as the frequency with which the absolute mean difference exceeds the actually observed value. If this 
$p$-value is smaller than the chosen significance level, accept 
${\mathrm H}_1$	
Otherwise, accept ${\mathrm H}_0$	
\end{enumerate}

We set the significance level to 0.01. For $\alpha$-helix proportion, the difference in mean values between the Robust and Random proteins in the original data is $39.18 - 33.16 = 6.02$ (see Table 1). Therefore, we accept 
${\mathrm H_1}$ if this observed difference falls within the upper 1\% tail of the bootstrap-based sampling distribution. For $\beta$-sheet proportion, the corresponding difference is $16.79 - 20.98 = - 4.19$; similarly, we accept ${\mathrm H_1}$ if this value lies in the lower 1\% tail of the bootstrap distribution.
In both cases, the observed mean differences indeed fell within the respective 1\% critical regions, leading to the acceptance of ${\mathrm H_1}$. More precisely, the $p$-value was zero in both analyses
---no single value from the resampled mean differences exceeded (or fell below, in the case of the negative difference) 
the actually observed mean difference.

\bibliography{main_revise2}
\bibliographystyle{unsrt}



\end{document}